\documentclass[useAMS,usenatbib]{mn2e}
\usepackage{amsmath}
\usepackage{graphicx}
\usepackage{color}
\title[The IMF and SFL in the outer disc of NGC 2915]{The initial mass function and star formation law in the outer disc of NGC 2915}

\author[S. M. Bruzzese et al.]{S. M. Bruzzese$^{1}$, G. R. Meurer$^{1}$, C. D. P. Lagos$^{2}$, E. C. Elson$^{3}$, J. K. Werk$^{4}$,
\newauthor John P. Blakeslee$^{5}$, and H. Ford$^{6}$\\
$^{1}$International Centre for Radio Astronomy Research, University of Western Australia, Crawley, WA, Australia\\
$^{2}$European Southern Observatory, Garching, Germany\\
$^{3}$Astrophysics, Cosmology and Gravity Centre, University of Cape Town, Rondebosch, South Africa\\
$^{4}$UCO/Lick Observatory, University of California, Santa Cruz, CA, USA\\
$^{5}$Herzberg Institute of Astrophysics, National Research Council of Canada, Victoria, BC, Canada\\
$^{6}$Department of Physics and Astronomy, Johns Hopkins University, Baltimore, MD, USA\\
}

\newcommand{\HI}{\mbox{H\,{\sc i}}}
\newcommand{\SFI}{{$\rm \Sigma_{SFR}$}}
\newcommand{\Ha}{{H$\rm \alpha$}}
\newcommand{\HII}{\mbox{H\,{\sc ii}}}
\newcommand{\Htwo}{H{$_2$}}

\begin{document}

\date{\today}

\pagerange{\pageref{firstpage}--\pageref{lastpage}} \pubyear{2014}

\maketitle

\label{firstpage}

\begin{abstract}
Using \textit{Hubble Space Telescope} (\textit{HST}) ACS/WFC data we present the photometry and spatial distribution of resolved stellar populations in the outskirts of NGC 2915, a blue compact dwarf with an extended \HI\ disc. These observations reveal an elliptical distribution of red giant branch stars, and a clumpy distribution of main-sequence stars that correlate with the \HI\ gas distribution. We constrain the upper-end initial mass function (IMF) and determine the star formation law (SFL) in this field, using the observed main-sequence stars and an assumed constant star formation rate. Previously published \Ha\ observations of the field, which show one faint \HII\ region, are used to provide further constraints on the IMF. We find that the main-sequence luminosity function analysis alone results in a best-fitting IMF with a power-law slope $\alpha=-2.85$ and upper-mass limit $\rm M_u = 60\,M_{\sun}$. However, if we assume that all \Ha\ emission is confined to \HII\ regions then the upper-mass limit is restricted to $\rm M_u \la 20\, M_{\sun}$. For the luminosity function fit to be correct we have to discount the \Ha\ observations implying significant diffuse ionized gas or escaping ionizing photons. Combining the \textit{HST} photometry with \HI\ imaging we find the SFL has a power law index $\rm N=1.53 \pm 0.21$. Applying these results to the entire outer \HI\ disc indicates that it contributes 11--28\% of the total recent star formation in NGC 2915, depending on whether the IMF is constant within the disc or varies from the centre to the outer region.
\end{abstract}


\section{Introduction}
\label{sec:intro}
Various empirical relations are used as prescriptions for star formation, allowing us to model galaxy evolution. Two of these relations are the stellar initial mass function (IMF), and star formation law (SFL). The IMF affects many observable properties of galaxies; with any given IMF one can estimate: the stellar mass of galaxies from luminosities \citep[e.g.][]{Bell:2003hs, Panter:2007gb, Baldry:2008hm}, the star formation rate (SFR) from \Ha\ emission \citep[e.g.][]{Kennicutt:1989cu, Kennicutt:1998id}, and star formation history (SFH) from colour-magnitude diagrams \citep[e.g.][]{Gallart:1996kn, Dolphin:1997cm, Dolphin:2000ft, Skillman:2003gh, Dolphin:2002db, Williams:2008ic, Weisz:2013vt}. Beyond individual galaxies, a well defined IMF is required to derive the cosmic SFH of the Universe \citep{Madau:1996tu,Lilly:1996gba}, and to model the formation and evolution of galaxies.  On the other hand, the SFL describes the relationship between the interstellar medium (ISM) and the observed SFR, allowing us to numerically model star formation, place constraints on these models, and determine which theories of star formation best match observations.

\subsection{Initial mass function}
The IMF was originally parametrized as a power law mass distribution $\xi(m) =dN/dm \propto m^{\alpha}$, with $\alpha \approx -2.35$ based on observed star counts in the Solar neighbourhood \citep{Salpeter:1955hz}, and later found to turn over at low masses \citep[$m \la 1$ M$_{\sun}$, ][]{Kroupa:2001ki,Chabrier:2003ki}. Theoretically the IMF should change with star-forming conditions \citep{Bate:2005cr, Larson:2005gm, Elmegreen:2004jk}, yet the observed IMF in star clusters and associations does not vary significantly in different regions of the Milky Way (MW) and Magellanic Clouds (MCs).  Any deviations observed in these regions are within the limits of statistical uncertainties \citep{Bastian:2010ig}, leading to the assumption of a universal IMF \citep{Scalo:1986wt,Kroupa:2001ki,Kroupa:2002cm,Chabrier:2003ki}.

As it is a common assumption that the IMF is universal, most studies in resolved stellar populations aim to solve for the SFH, while assuming an IMF, by analysing colour-magnitude diagrams (CMDs) \citep[e.g.][]{Gallart:1996kn,Aparicio:1997iv,Barker:2007vj,Williams:2008ic,Williams:2009hk, Williams:2010hw, Lianou:2013ej,Harris:2009bz, Weisz:2011hn, Weisz:2013vt}. These studies generally assume a \citet{Salpeter:1955hz} or \citet{Kroupa:2001ki} IMF with an upper mass limit $\rm M_u = 100-120\, M_{\sun}$. In this paper we consider a \citet{Kroupa:2001ki} IMF with $\rm M_u=120\, M_{\sun}$ and $\alpha = -2.35$ to be standard and will refer to it as the Kroupa IMF. However, recent observational results \citep{Hoversten:2008ku,Meurer:2009gp, Lee:2009bs, Gunawardhana:2011hm} based on the integrated light of  galaxies indicate that the upper end of the IMF may vary with galactic luminosity, surface brightness or SFR, challenging the assumption of a universal IMF. If the IMF is not universal and has steeper slope, or the upper mass limit is lower than the assumed IMF, then a recent SFH that is truncated or declining could be inferred when the SFR is constant \citep{Meurer:2009gp}. 

In this study rather than solving for the SFH we explore the less well used approach of adopting a plausible SFH and solving for the IMF. Using this method to determine the IMF in extreme environments may provide insight into IMF variations. One such extreme environment is the outer disc region of gas-rich star-forming galaxies. These regions have very low gas and stellar-surface densities and have recently been shown to harbour low-intensity star formation \citep{Thilker:2007ho}. Our method is justified by the relative simplicity and quiescence of the outer disc on time-scales shorter than the dynamical time. Since it is thought the IMF may be deficient in high-mass stars in low density environments \citep{Elmegreen:2004jk}, studying the IMF in the outer discs of gas-rich, star-forming galaxies may provide insight as to how the IMF depends on environment.  

\subsection{Star formation law}
The empirical relation between star formation and gas density was introduced by \citet{Schmidt:1959bp}, who proposed that the rate of star formation varies with a power of the gas density $\rho_{SFR} \sim \rho_{gas}^n$. In this study \citet{Schmidt:1959bp} compared the spatial distribution of young stars as tracers of recent star formation to \HI\ gas in the MilkyWay, finding $n=1-3$. Studies of the spatially resolved or local SFL in the nearest galaxies followed, using the surface densities of OB stars and \HII\ regions as tracers of recent star formation and \HI\ observations to trace the gas \citep[e.g.][]{Sanduleak:1969hn, Madore:1974bg, Hamajima:1975uh}. These studies used projected gas surface densities, in which case the SFL can be parameterized as $\Sigma_{SFR}=A \Sigma_{gas}^N$ and found $N=1.1-3.1$. Note that for a constant scale height $n=N$, thus the exponent is the same for both volume and surface densities. Later studies of the local SFL started to incorporated the \Htwo\ content estimated from CO observations in the $\Sigma_{gas}$ estimates; this included \citet{Kennicutt:1989cu} who found $N = 1.3\pm 0.3$. Rather than studying the SFL on local or sub-galactic scales \citet{Kennicutt:1998id} determined a global SFL by studying the disc averaged properties of an extensive sample of galaxies. \citet{Kennicutt:1998id} derived a global SFL power-law index of $N = 1.4 \pm 0.15$ and $A = (2.5 \pm 0.7) \times 10^{-4} \, \rm M_{\sun}\, yr^{-1}\, kpc^{-2}$, where $\Sigma_{SFR}$ is the star formation intensity derived from \Ha\ emission and $\Sigma_{gas}$ is the total gas (\HI\ + \Htwo) mass surface density. To a large extent this has become the `standard' SFL, bearing in mind that the scale factor depends on the adopted IMF, and an assumed constant SFR over the short life time of O stars ($\la$ 10 Myr)\footnote{The quoted $A$ is for a Salpeter IMF over the mass range 0.1 to 100 $\rm M_{\sun}$, whereas for a Kroupa IMF it is a factor of 1.7 lower \citep{Salim:2007hq}.}. Recent studies of the local SFL go down to sub-kpc resolution and find that $N$ takes on a wide range of values from super-linear ($N>1$) \citep{Boissier:2003hn, Kennicutt:2007dz, Genzel:2013iq} and linear ($N=1$) \citep{Bigiel:2008bs}, to sub-linear ($N<1$) \citep{Shetty:2013bp}.

\begin{figure*}
\centering
\includegraphics[width=130mm, height=85mm]{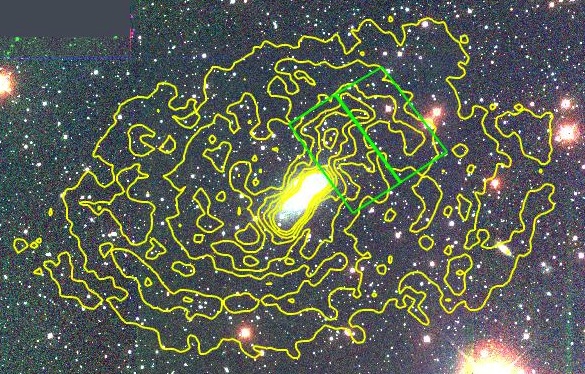}
\caption{Three colour image of NGC 2915 from the Cerro Tololo Inter-American Observatory 1.5-m telescope using \textit{V}-band (blue), \textit{R}-band (green), and \textit{I}-band (red)  filters. The ACS/WFC footprint is shown in green, covering the observed outer-disc region of NGC 2915. \HI\ data from the Australian Compact Telescope Array and are shown as yellow contours indicating \HI\ surface mass densities of: 1.4, 2.7, 3.9, 5.2, 6.4, 7.6, 8.9 $\rm M_{\sun}\,{pc}^{-2}$ \citep{Elson:2010iv}.}
\label{fig:ACS_HI}
\end{figure*}

There has been much discussion as to what phase of the ISM the SFL pertains, greatly affecting the derived value of $N$. Initially the SFL was calibrated with \HI\ + \Htwo\ (with \Htwo\ traced by CO), yielding super-linear values of $N$ \citep{Kennicutt:1989cu, Kennicutt:1998id}. \citet{Kennicutt:1989cu} showed that on global scales the SFR is more strongly coupled to the total gas density than either \HI\ or \Htwo\ alone. However, recent work shows that on smaller scales, and in the optically bright part of galaxies, the SFR scales better with \Htwo\ and not at all with \HI; yielding lower values of $N$ and a nearly linear SFL \citep{Bigiel:2008bs, Leroy:2008jk, Schruba:2011em}. In the outer regions of galaxies the SFR correlates with \HI\ surface densities \citep{Bigiel:2010jc, Bolatto:2011et}, but this could be consistent with star formation correlating with \Htwo, which largely remains undetectable \citep{Krumholz:2013kk}.

Understanding how the SFL depends on environment is crucial to determine the underlying physics of star formation. The observed scatter in composite SFLs and variation of the slope between different galaxies is an indication that the SFL is not solely dependent on the surface density of gas, but varies with environment. This idea has been supported by observations and theoretical models \citep{Schruba:2011em, Leroy:2008jk, Dopita:1994cm}. In addition, a large scatter is observed in the correlation between SFR and \HI\ surface density in \HI\ dominated regions, as shown by \citet{Bigiel:2010jc}. The observed scatter may be explained by recent models of  \citet{Ostriker:2010dm} and \citet{Krumholz:2013kk}, which show that the SFR in the low density \HI\ dominated regime should depend sensitively on parameters such as metallicity and the density of stars plus dark matter. Thus, the low-density outer discs of gas-rich star-forming galaxies provide a unique environment to determine how well these models match observations. 

\subsection{This work}

In this paper we determine the IMF and the SFL in the outer \HI\ disc of the dark matter dominated galaxy NGC 2915. This galaxy has been labelled the `Ghost Galaxy'; when viewed in optical light it is classified as a blue compact dwarf (BCD) \citep{Meurer:1994kv} but when observed in \HI\ it appears to be a late-type spiral extending over five times its Holmberg radius \citep{Meurer:1996ca, Elson:2010iv}. Beyond its core of intense star formation, this galaxy showed no signs of extended star formation in earlier studies \citep{Meurer:1994kv, Meurer:1996ca}. However, \Ha\ images published by \citet{Werk:2010fn} show three very faint \HII\ regions in the outer disc, indicating that some high-mass star formation is occurring. The \textit{HST}/ACS images presented by \citet{Meurer:2003fz} were intended to find young stars associated with star formation in the outer \HI\ disc. These observations as well as the data from \citet{Werk:2010fn} and \citet{Elson:2010iv} are combined in this study to determine the SFL and place constraints on the upper-end of the IMF.  We study NGC 2915 because of its high mass-to-light ratio, low stellar-density, and low intensity star formation, which make it an ideal target for understanding these two empirical relations in an extreme environment.

This paper is organized as follows: In \S \ref{sec:obs} we present  \textit{HST}/ACS observations and photometry, including photometric errors and completeness derived from artificial star tests. In \S \ref{sec:CMD} we present the CMDs and stellar content of NGC 2915, including the spatial distribution of stars in different evolutionary stages. In \S\ref{sec:IMF} we outline our technique for simulating CMDs, our statistical analysis to determine the best-fitting IMF parameter, and present our results. We do this first using the MS luminosity function analysis, and then applying the \Ha\ observations. In \S\ref{sec:SFlaw} we determine the SFL in the outer disc using the observed MS stars, and compare it to models of star formation. Finally, we present our conclusions in \S \ref{sec:con}.

\begin{table}
\centering
\caption{Adopted NGC 2915 properties}
\label{tab:info}
\begin{tabular}{|l|l|l|}
\hline 
\hline
Property & Value  & Reference\\ 
\hline 
Z & 0.4 Z$_{\sun}$ & \citet{Werk:2010fn} \\

$E(B-V)_{\rm tot}$ & 0.45 mag & \citet{Werk:2010fn} \\ 

$D$ & $4.1 \pm 0.3$ Mpc & \citet{Meurer:2003fz} \\ 

$r_H$ & 2.3 kpc & \citet{Meurer:2003fz} \\

$i$ & $55^{\circ}$ & \citet{Elson:2010iv} \\ 

$\rm M_{\star}$ & $3.2 \times 10^{8}\,\rm M_{\sun}$ &  \citet{Meurer:2003fz} \\

$\rm M_{dyn}$ & $151.6 \times 10^{8}\,\rm M_{\sun}$ &  \citet{Elson:2010iv} \\

$\rm M_{dyn}/L_{B}$ & $140.91\, \rm M_{\sun}/L_{\sun,B}$  & \citet{Elson:2010iv}\\

$\rm M_{\rm{H}I}/L_{B}$ & $6.27\, \rm M_{\sun}/L_{\sun,B}$ & \citet{Elson:2010iv}\\ 
\hline 
\end{tabular} 
\end{table}

\section{Observations and photometry}
\label{sec:obs}
\begin{figure}
\centering
\includegraphics[width=85mm, height=85mm]{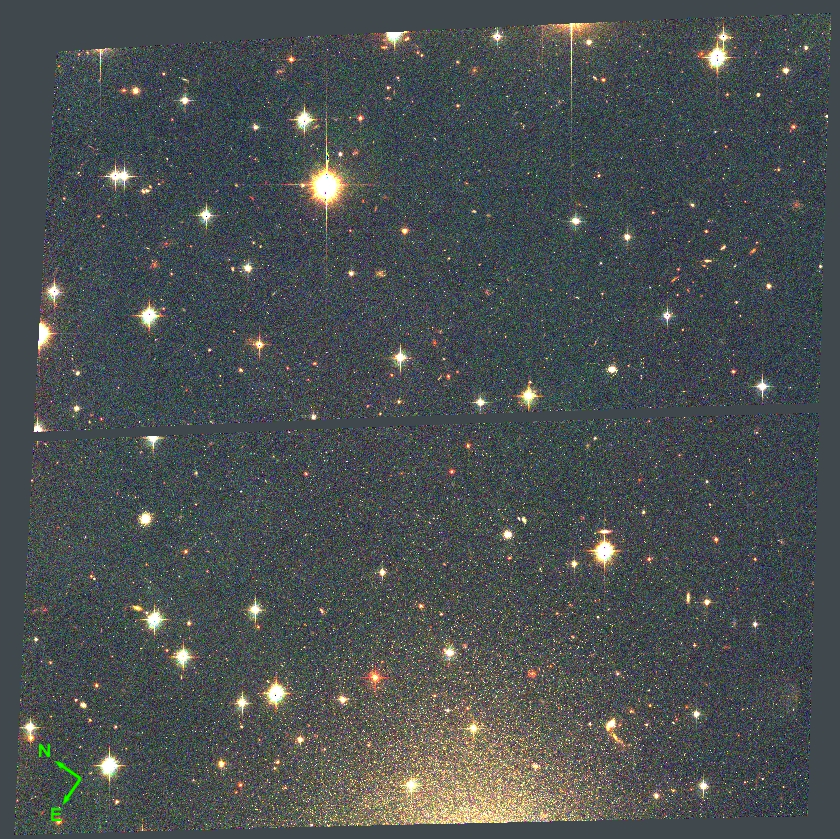}
\caption{Three colour image of NGC 2915 from ACS/WFC data in F475W (blue), F606W (green), and F814W (red) filters. The WFC field of view is 202 $\times$ 202 arcsec which corresponds to 4015 $\times$ 4015 pc. The centre of NGC 2915 is located below the bottom of the image. The orientation of the image is given by the green compass at the lower left.}
\label{fig:RGB}
\end{figure}

The primary data used in this study were taken by the Advanced Camera for Surveys (ACS) Wide Field Camera (WFC) on board the \textit{HST} (proposal ID: 9288) in December 2002. Four exposures were obtained each in the F475W (\textit{g}-band), F606W (\textit{V}-band), and F814W (\textit{I}-band) filters, with total exposure times of 2480, 2600, and 5220 seconds respectively. These images are centered on $\rmn{R.A.}(J2000.0)=09^{\rmn{h}} 25^{\rmn{m}} 36\fs48$ and $\rmn{Decl.}(J2000.0)=-76\degr 35\arcmin 52\farcs4$, covering radii 45 -- 257 arcsec (0.89 -- 5.12 kpc) from the centre of NGC 2915. This position was designed to contain both the strongest \HI\ concentration outside of the core and one of the three \HII\ regions \citep[subsequently discussed by][]{Werk:2010fn}, but otherwise to avoid the core\footnote{These observations were also designed to allow a second outer-disc \HII\ region to be imaged by the High Resolution Camera (HRC). However, due to an adjustment to the aperture definitions, shortly before the observations, the parallel HRC observations missed their target.}. This was done in order to maximize the possibility of detecting high-mass stars. Fig. \ref{fig:ACS_HI} shows the location of the ACS/WFC pointing in relation to the rest of NGC 2915, and Fig. \ref{fig:RGB} shows a colour version of the drizzled ACS/WFC images. Table \ref{tab:info} lists the physical properties of NGC 2915 used throughout this paper.

\subsection{Data reduction and measurement}
\label{sec:phot}
Basic image processing was done using \textsc{Calacs} \citep{Hack:1999wl}, which produced calibrated and combined cosmic ray rejected (CRJ) images. \textsc{Apsis} \citep{Blakeslee:2003wy} was used to align and combine the CRJ images to produce single, geometrically corrected drizzled images in each filter with a pixel size of 0.05 arcsec (0.99 pc), which we use as a reference image in the point source photometry described below.

We performed stellar point source photometry using the ACS module of the stellar photometry package \textsc{Dolphot} (v 2.0), a modified version of \textsc{HSTphot} \citep{Dolphin:2000ib}, designed specifically for resolved stellar photometry of ACS data. \textsc{Dolphot} computes the photometry of all stars individually using pre-computed point spread functions (PSFs) for each filter. We selected a drizzled image in the F606W filter as the initial detection and position reference image, and performed photometry simultaneously on all CRJ images, in all filters. We followed the processing steps outlined in the \textsc{Dolphot/ACS} User's Guide, with minor modifications to handle \textsc{Apsis} products. We adopted \textsc{dolphot} parameters similar to those used by the ACS Nearby Galaxy Treasury (ANGST) team \citep{Dalcanton:2009iy}. They found \textit{Force1}, \textit{RAper} and \textit{FitSky} to have the strongest influence on photometry; we have set these parameters to the values suggested.

\textsc{Dolphot} provides photometry, position, and quality parameters for each detected star. All photometry is given in the VEGAMAG system using transformations from \citet{Sirianni:2005it}, and corrected for CTE loss according to \citet{Riess:2004wt}. To convert photometry to the ABMAG system one should add -0.100, 0.257, 1.275 mag respectively to the $g, V, I$ photometry presented here \citep{Sirianni:2005it}.  In order to select objects likely to be stars from the photometric output we made the following quality cuts: $(S/N)_{1,2}> 4$; $|sharp_{1} + sharp_{2}| < 0.05$; $|round_{1}  + round_{2}| < 1.4$; and $(crowd_{1} + crowd_{2}) <0.6$. These cuts were done separately with the \textit{g,V} bands or \textit{V,I} bands representing filters 1,2. The star lists of the separate cuts were merged so that the survival in one set of cuts was sufficient for inclusion in our final catalogue. The sharpness and roundness cuts were chosen to remove sources likely to be diffraction spikes or cosmic rays along with blended stars and background galaxies. The crowding cuts were chosen to remove stars with photometry  significantly affected by crowding. These quality cuts produced the cleanest CMDs, minimising false stellar detections from background galaxies, extended sources, and saturated pixels. The final photometric catalogue contains 27156 stellar detections ($\sim10\%$ of the original \textsc{dolphot} output).

\begin{figure}
\centering
\includegraphics[width=85mm, height=85mm]{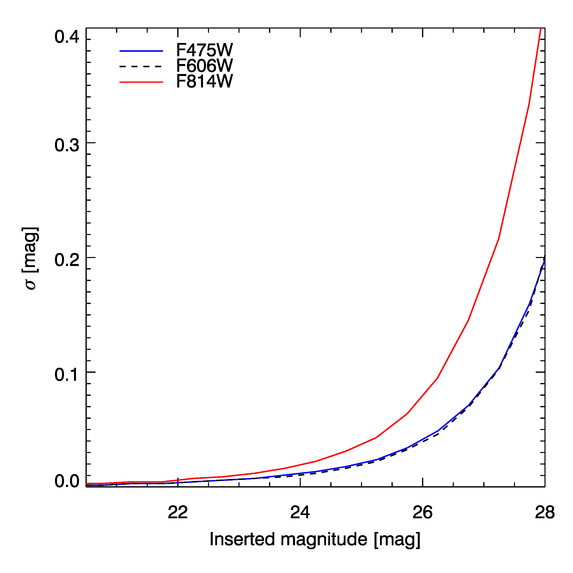}
% scat_gv.pro
\caption{Median photometric errors derived from artificial star tests.}
\label{fig:phot_err}
\end{figure}

\begin{figure*}
\centering
%mag_hess.pro
\includegraphics[width=85mm, height=115mm]{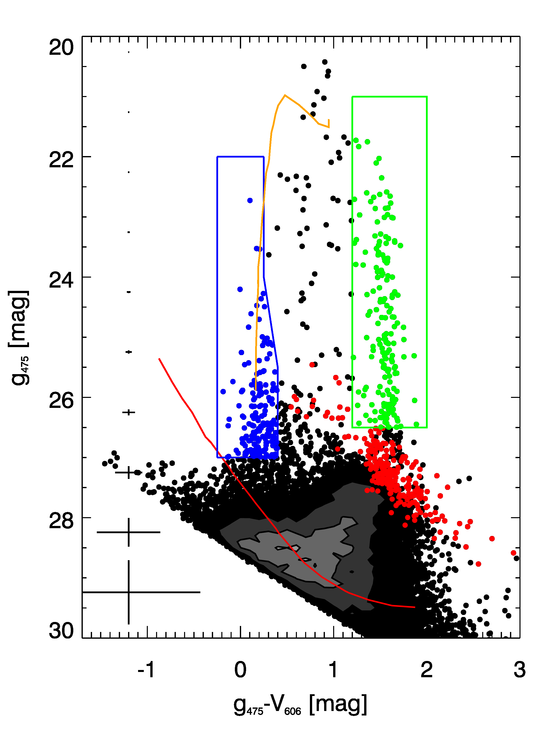}
\includegraphics[width=85mm, height=115mm]{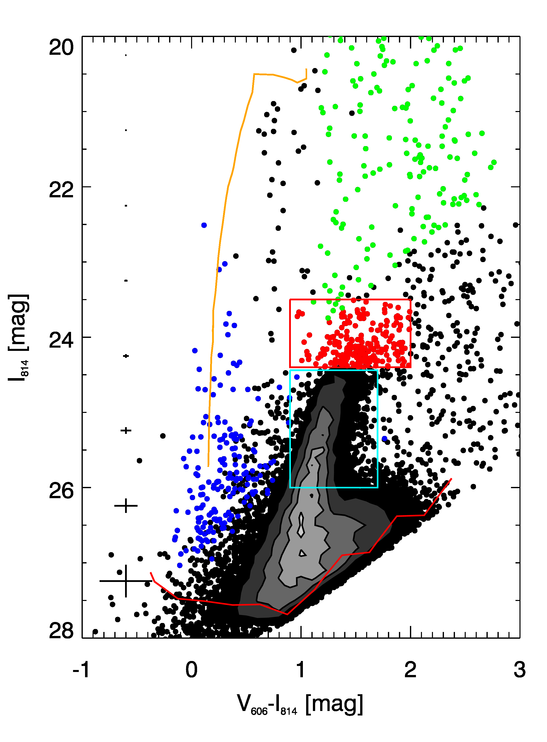}
\caption{CMDs of the ACS/WFC data in $g$ versus $(g-V)$ (left) and $I$ versus $(V-I)$ (right). Stars meeting our stellar quality cuts are shown as dots with a Hess diagram over-plotted with contours of $2.5, 5, 10, 15, 20 \times 10^{3}$ stars mag$^{-2}$. Uncertainties derived from artificial star tests are shown as error bars on the left of each CMD and the 60\% completeness limits are shown as red lines. Polygons are used to identify different stellar evolutionary phases. \textbf{\textit{(g-V)} CMD}: The blue polygon defines the MS stars, and the green rectangle defines the foreground stars. \textbf{\textit{(V-I)} CMD}: Cyan rectangle defines RGB stars and the red rectangle defines AGB stars. The Geneva stellar evolutionary track for a $25\, \rm M_{\sun}$ star is shown in orange for reference.}
\label{fig:CMD}
\end{figure*}

\begin{figure*}
\centering
% colour_mag_xy.pro
\includegraphics[width=175mm, height=175mm]{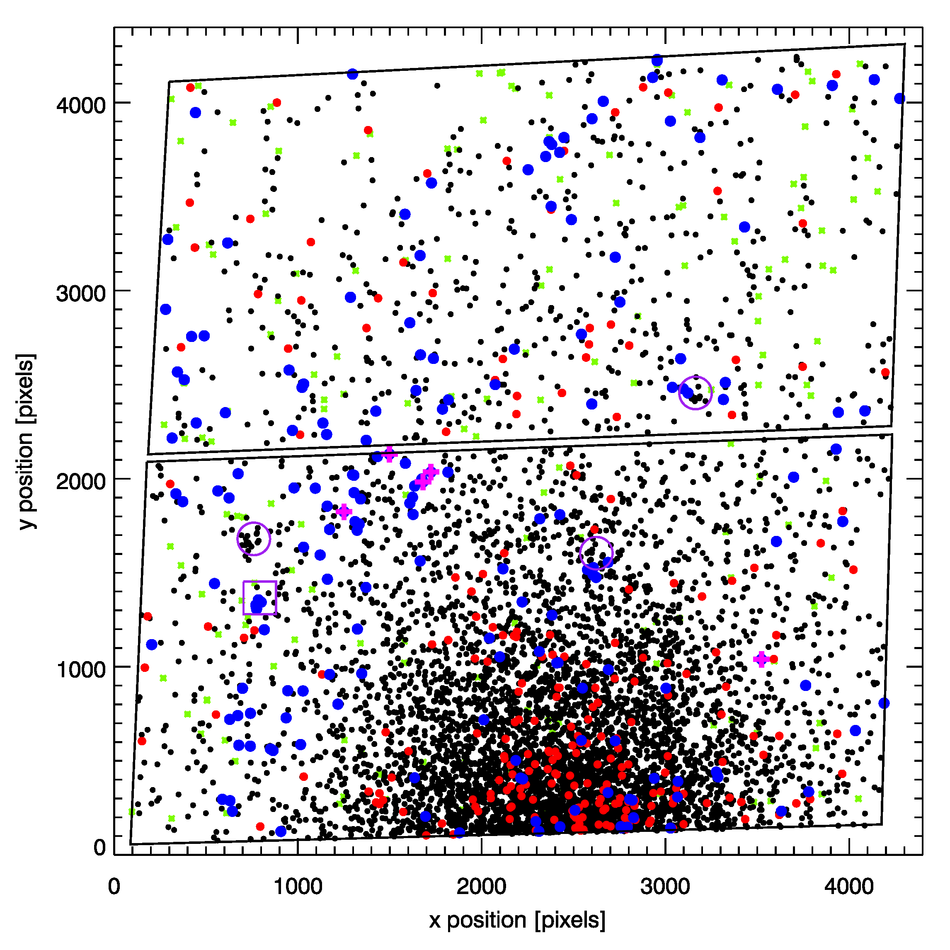}
\caption{Spatial distribution of stars in NGC 2915 as identified by the regions shown in the CMDs in Fig. \ref{fig:CMD}. RGB stars are identified as black points, AGB stars as red points, MS stars as blue circles, and foreground stars as green asterisks. The five most optically luminous MS stars are identified as magenta crosses. The RGB and AGB stars follow an elliptical distribution centred beyond the bottom of the image, the blue MS stars have a clumpy distribution, and the foreground stare are evenly distributed across the image consistent with stellar foreground contamination. Large purple circles indicate the location of the three globular clusters discovered by \citet{Meurer:2003fz}. The large purple square shows the location of H{\sc ii}-1 from \citet{Werk:2010fn}. The black outline shows the regions covered by the WFC CCDs.}
\label{fig:spat_dist}
\end{figure*}

\subsection{Artificial star tests}
We determine completeness and photometric errors using artificial star tests in \textsc{Dolphot} \citep[similar to][]{Dalcanton:2012ic}. This is done to determine how well stars of known colour and magnitude are recovered with \textsc{Dolphot}. We simulated a total of $3 \times10^{5}$ stars, distributed evenly across the images. The artificial stars have colours and magnitudes similar to the observed data but extended in range by $\sim1$ mag to allow for shifts due to crowding and errors (e.g. $19<g<30$ and $1<g-V<3$). We apply the same photometric cuts that were applied to the observed data to produce a catalogue of found artificial stars. Fig. \ref{fig:phot_err} shows the median error as a function of apparent magnitude, computed using Gaussian statistics,  as the median absolute deviation between the inserted and recovered magnitudes, for all recovered artificial stars. The completeness is the percentage of recovered stars compared to inserted stars in binned regions (0.25 in colour and 0.5 in magnitude) of the CMD. The 60\% completeness limit for each CMD is shown in Fig. \ref{fig:CMD}. These tests demonstrate that the main features of the CMD suffer at most minor incompleteness, which we correct for in our simulations (see \S\ \ref{sec:sim_CMD}).

\section{Stellar content}
\label{sec:CMD}
\subsection{Colour magnitude diagrams}
The CMDs shown in Fig. \ref{fig:CMD} reveal several features common to composite stellar populations \citep[e.g.][]{Gallart:2005ws}. These include a prominent red-giant branch (RGB), MS, and asymptotic-giant branch (AGB). The RGB stars indicates star formation occurring $>$ 1 Gyr ago, the high-mass MS stars indicates recent star formation ($ \la 100$ Myr), and the presence of AGB stars indicates stars formation between 1 and 10 Gyr ago.  Fig. \ref{fig:CMD} may also show a population of blue loop stars, which formed between 30 and 100 Myr ago. Blue loop stars are high-mass stars that cross back towards the blue region of the CMD post hydrogen-burning. Although their numbers are small, their colours are similar to MS stars \citep{Tosi:1991ed, DohmPalmer:1997jr, Dalcanton:2012ic}, hence they may contaminate our selection of MS stars, so we take pains to account for them in our analysis.

Fig. \ref{fig:CMD} shows the selection boxes we use to select different stellar populations. In Fig. \ref{fig:spat_dist} we plot the spatial distribution of MS, RGB, AGB, and bright foreground stars, as identified from their positions on the CMDs. The red stars (RGB and AGB) have a smooth elliptical  distribution centered below the bottom of the image. This extended distribution suggests that the optical light from NGC 2915 is dominated by old population {\sc ii} stars; a view supported by the very red colour $(B-R)_0=1.65$ found for $r \ga 0.8$ kpc by \citet{Meurer:1994kv}, and the discovery of three globular clusters \citep{Meurer:2003fz}, which we have plotted in Fig. \ref{fig:spat_dist}. This illustrates that BCDs are typically old systems undergoing an intense episode of recent star formation in their core \citep[e.g.][]{Thuan:1983eo, Aloisi:2007cd, Zhao:2011ko}. Apart from a slight enhancement of RGB stars near two of the globular clusters, the distribution of red stars is very smooth. In contrast, the young, blue MS stars have a clumpy distribution with a much lower density compared to the red stellar populations. As we show in $\S$ \ref{sec:SFlaw} their density is correlated with the \HI\ emission. These stars indicate recent ($\la$160 Myr) star formation in the outer disc of NGC 2915. The likely foreground stars are evenly distributed over the entire image as expected. We modelled stellar foreground contamination using the population synthesis code \textit{TRILEGAL} \citep{Girardi:2005bb}, assuming constant foreground galactic extinction, a Kroupa IMF with no binaries, and using the standard inputs for the MW disc. These simulations indicate that we expect one foreground star in our MS selection box, which we consider negligible.

\subsection{Ionizing stars and the most luminous stars}
\label{sec:HII}
\begin{figure}
\centering
\includegraphics[width=85mm, height=109mm]{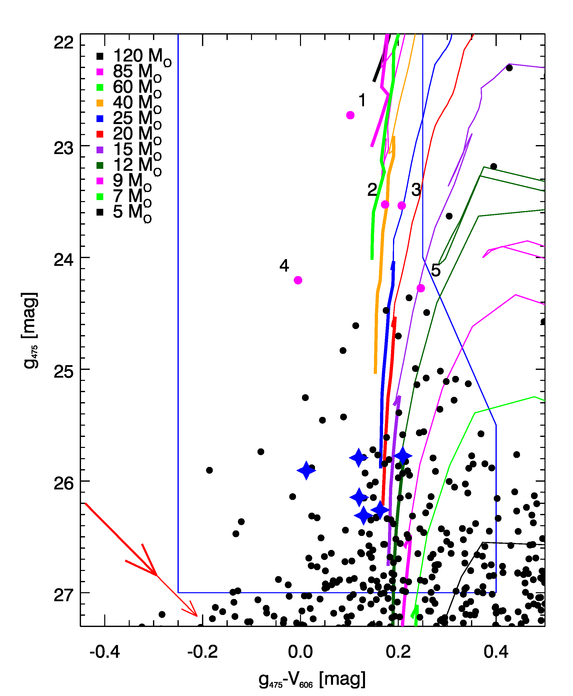}
% massive_plot.pro
\caption{Zoomed in version of the $g$ versus $(g-V)$ CMD MS selection polygon, which is outlined in blue. The six \HII-1 stars, indicated in Fig. \ref{fig:zoom_HII}, are marked as blue stars here. The five most luminous MS stars identified in Fig. \ref{fig:massive_5}, are shown in magenta and labelled. The high-mass ($\rm 5\,M_{\sun}<M<120\,M_{\sun}$) Geneva evolutionary tracks for $Z=0.008$ \citep{Schaerer:1993wj} are shown as coloured lines (with the mass legend in the upper-left); the hydrogen-burning phases are indicated by thicker tracks. The internal, $E(B-V)_{i}=0.175$ (red), and foreground $E(B-V)_{f}=0.275$ (bold red), extinction applied to the evolutionary tracks are shown by the arrows. The MS selection box is narrower at higher luminosities to avoid contamination from blue-loop stars.}
\label{fig:HII_reg}
\end{figure}

\begin{figure}
\centering
\includegraphics[width=65mm, height=65mm]{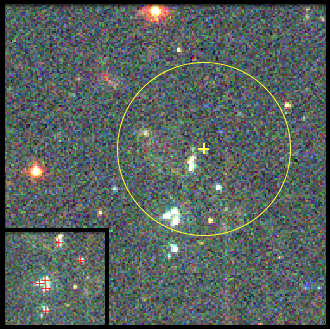}
\caption{Three colour ACS/WFC image showing in detail the stars associated with \HII-1 from \citet{Werk:2010fn}. The `+' marks the position given by \citet{Werk:2010fn}. The yellow circle centred on this position has a 2.5 arcsec (50 pc) radius indicating the expected combined effects of seeing (2.3 arcsec) in the images of \citet{Werk:2010fn}, and the expected astrometric uncertainty ($\sim$ 1 arcsec) between the \textit{HST} pointing and the work of \citet{Werk:2010fn}. The inserted image in the bottom left corner shows the six stars that pass the stellar classification criteria, marked by red crosses.}
\label{fig:zoom_HII}
\end{figure}

Our field contains H {\sc ii} region 1 (hereafter  H {\sc ii}-1), discussed in detail in \citet{Werk:2010fn}. Its location is marked in Fig. \ref{fig:spat_dist} by the purple box. It corresponds to a loose association of MS stars, which we highlight in the $(g-V)$ CMD in Fig. \ref{fig:HII_reg}. A zoomed in image (taken from Fig. \ref{fig:RGB}) of the loose association is shown in Fig. \ref{fig:zoom_HII}. The objects marked as stars, in this region, are very close together or have an elongated appearance indicating there are neighbours that have not been resolved with these data and software. These stars typically have a greenish appearance in our $gVI$ three colour images, and are embedded in a low surface-brightness green emission. This colour may be due to contamination by emission lines such as  H $\rm \beta$, [{\sc Oiii}] $\lambda\lambda$4959,5007\AA, \Ha, and [{\sc Sii}] $\lambda\lambda$6716,6731\AA, all of which fall in the passband of the F606W $V$-band filter.  Six of the stars in H {\sc ii}-1 make the quality cuts applied to the photometry (see $\S$\ref{sec:phot}), all of which occupy the CMD in the lower luminosity region of our MS selection polygon ($26.4<g<25.7$). We identify these stars as a loose association with a projected diameter $<25$ pc, similar in size to those found in the LMC by  \citet{Lucke:1970kf}.

\HII\ regions indicate the presence of young, high-mass ($\rm M_{\star}>15\,M_{\sun}$), O-type stars which produce UV radiation capable of ionizing the surrounding ISM. In Fig. \ref{fig:HII_reg} we show the stellar evolutionary tracks used in our modelling, which demonstrate that the stars we detect are consistent with having $\rm M_{\star}>15\,M_{\sun}$. \citet{Werk:2010fn} measured a total \Ha\ luminosity of $1.6 \times 10^{36} \, \rm erg\, s^{-1}$ for H {\sc ii}-1, or an ionizing photon rate of $1.1 \times 10^{48}\rm \, s^{-1}$. Here we make the standard assumption that all ionizing photons are captured by the ISM near the O stars to produce discrete \HII\ regions, i.e. case B recombination. The ionizing photon rates from \citet[][table 1]{Martins:2005bn}, show that a single luminosity class {\sc v} star (i.e. MS) with spectral type between O8.5 and O9 can produce the ionizing flux observed in H {\sc ii}-1. Hence, only one of the six stars in the association needs to be the ionizing source, with the rest being non-ionizing. Alternatively the ionizing flux may be comprised of the output of several lower mass early type B stars \citep[see table 3.1 of][]{Conti:2008ur}

\begin{figure*}
\centering
\includegraphics[width=170mm, height=57mm]{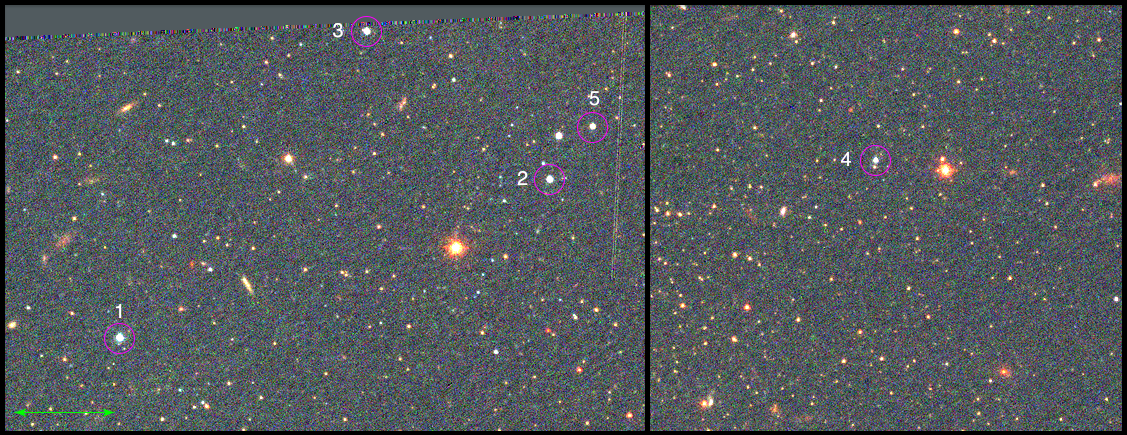}
\caption{Three colour ACS/WFC images showing in detail the five most luminous MS stars, labelled in order of descending luminosity. The left panel shows the four other most luminous stars from our data; these stars are located left of the middle near the top of WFC1. The right panel shows the fourth brightest MS star located in the bottom right corner of the lower WFC CCD chip (WFC1). The location of these five stars with respect to the rest of the field can be seen in Fig. \ref{fig:spat_dist} and Fig. \ref{fig:HI_dist} as magenta points. The green scale bar located in the bottom left corner is 5 arcsec (99 pc).}
\label{fig:massive_5}
\end{figure*}

Naively one expects the ionizing stars to be the most luminous stars in our MS selection box. However, Fig. \ref{fig:HII_reg} shows that of the six possible ionizing stars, none are the most luminous in the optical. In Fig. \ref{fig:HII_reg} and Fig. \ref{fig:massive_5} we identify the most luminous stars in the MS selection box, numbering them by decreasing $g$-band luminosity. Fig. \ref{fig:massive_5} shows zoomed in images of the most luminous stars. Four of them (numbers 1,2,3, and 5) are located within a circle of projected radius $\sim 240$ pc near the top-centre of  WFC1, while the other is located near the bottom-right corner of WFC1. All five are circular, consistent with being single sources, while none have the greenish tinge of the \HII-1 stars. The brightest source has three faint sources with 0.6 arcsec (12 pc), suggesting it is a compact group. The proximity of the four stars suggests they are part of a dissolving group or association. If it is spreading at a velocity equal to the \HI\ velocity dispersion then it would take 21 Myr for a very compact group to expand to the observed projected radius. Thus, we should not see stars with $\rm M_\star > 15\, M_{\sun}$, which have a MS lifetime of $\sim 13$ Myr, in such an expanding group. This is consistent with none of the stars producing ionizing radiation. While the evolutionary tracks in Fig. \ref{fig:HII_reg} suggest that the most luminous stars may have $\rm M_{\star} \approx 25-85\, M_{\sun}$, as noted above, it is possible that the brightest stars also include some multiples or post-MS stars, which we discuss further in \S\ \ref{sec:clust}.

\section{Constraints on the initial mass function}
\label{sec:IMF}

\subsection{Choice of SFH}
\label{sec:SFH} 

The IMF and SFH are degenerate; a lack of upper main sequence (MS) stars may be due to those stars not forming (IMF), or recent drop in the SFR (SFH). So in our approach of assuming the SFH, the choice of its functional form is crucial. The SFH of BCDs, such as NGC 2915, is typically assumed to be dominated by a short duration burst. Bursts of star formation are thought to occur due to a build up of gas in the central region of the galaxy, which triggers star formation once a critical density is achieved. \citet{McQuinn:2010db} used \textit{HST} CMD analyses of the starbursts in dwarf galaxies to determine that bursts last on the order of 0.5 to 1 Gyr time-scales. They found that earlier estimates of burst time-scales of a few Myr relate to only a small part of the galaxy \citep[e.g. star cluster, as suggested by][]{Meurer:2000ur}. Instead the new analysis is consistent with earlier work based on broadband colours, which yielded burst time-scales of $\ga$100 Myr for ongoing star formation \citep[e.g.][]{Meurer:1992kb, Marlowe:1999jd}. These time-scales are also consistent with causality arguments that bursts should not be shorter than the crossing or dynamical times.

While the inner part of NGC 2915 has a BCD morphology, here we are concerned with the outskirts, which have a much more quiescent appearance. NGC 2915 has a fairly symmetric disc with a \HI\ velocity dispersion of 8--12 km s$^{-1}$ for $r>$200 arcsec \citep{Elson:2010iv}. These properties indicate that there has been no recent interaction that would cause an increase or burst of star formation in the outer disc \citep{Meurer:1996ca, Werk:2010fn}. Since the blue stars in our images are spread throughout the field the minimum time-scale we can expect star formation to have lasted is the crossing time of the field. For a disturbance to traverse the ACS/WFC field of view, travelling at the \HI\ velocity dispersion, this is 480 Myr. The lower luminosity limit of our MS selection box corresponds to the peak luminosity of a $\sim 4\, \rm M_{\sun}$ star, which have a MS lifetime of 160 Myr. As long as star formation lasts this long or longer, it will effectively be equivalent to a constant SFR. Since the minimum expected time-scale is three times longer than this, we conclude that a constant SFR is a reasonable assumption. 

We also considered smoothly declining SFHs that are astrophysically based. First, we considered  a SFH declining like the Cosmic SFR density. Following \citet{Hopkins:2006bq} we estimate a 7\% decrease in the SFR from 160 Myr ago to now (z=0.01 to z=0). Second, we considered a SFH based on an alternative version of the SFL found in galaxies, which \citet{Kennicutt:1998id} determined to be equivalent to the conversion of 10\% of the available gas mass into stars each orbit. Using the rotation curve of \citet{Elson:2010iv} the average orbital time is $\sim230$ Myr for the radii probed by our ACS/WFC data. This corresponds to a 14\% decrease in the SFR over 160 Myr. We find the change in the SFR for both SFHs, over the 160 Myr time-scale that our analysis is sensitive to, to be negligible and produce results equivalent to a constant SFR. 

We note that it is not clear that a declining SFR is applicable in the outer disc of NGC 2915.  \citet{Elson:2011jq} find evidence in the velocity field that there is a net outward radial motion of the gas in the disc of NGC 2915. This may be due to conservation of angular momentum; as some gas is being funnelled in to the centre to feed the starburst, there is also a net outflow at large radii. Hence, the gas density at large radii can be increasing and likewise the local SFR. This is consistent with the disc spreading scenario of \citet{Roskar:2008fv}. These growth times should also be limited by the dynamical time, and be large compared to the lifetime of the stars we are analysing.

It is not just the temporal evolution of the SFR, but also the small scale spatial environment of star formation, which may be the key to understanding the nature of upper-end IMF variations. Indeed, two competing models of high-mass star formation have different predictions of where high-mass stars are born. In the competitive accretion model \citep[e.g.][]{Bonnell:2002et} high stellar masses are produced by the interactions of proto-stars in dense environments, which require the high densities of star clusters to form. In the monolithic collapse model \citep[e.g.][]{Yorke:2002dn} the final stellar mass depends on the properties of the molecular gas core, allowing high-mass stars to form in isolation. Following the review of  \citet{Lada:2003il} it has been common to assume that all high-mass star formation is confined to star clusters. However, in that work the definition of `clusters' is rather expansive and includes sources that are unbound upon clearing of their nascent ISM. These are not structures where competitive accretion should be effective.

Examination of the distribution of Galactic protostars shows that they form over a wide range of gas column densities and stellar densities and that there is no clear break in the continuum between cluster and field \citep{Gutermuth:2011he}. The IMF of protostars stars that form appears to be spatially varying, being more weighted towards high-mass stars in dense regions than in the field \citep{Hsu:2012kf}, which is broadly consistent with competitive accretion. On the other hand, \citet{Bressert:2012gt} showed compelling evidence for the formation of O stars in isolation in the field surrounding R136 in the LMC, which suggests that competitive accretion is not the only mechanism for high-mass star formation. In addition, \citet{Goddard:2010io} has shown that the cluster formation efficiency (CFE), the fraction of stars that form in clusters, decreases with star formation intensity. \citet{Kruijssen:2012bs}, has theoretically derived  the CFE as a function of galaxy properties and shown that the fraction of stars forming in clusters is $\sim 1$\% in low density environments, such as the outer disc of NGC 2915. Based on these results we model star formation in the outer disc of NGC 2915 as non-clustered, random sampling of the IMF. We revisit the possibility of clustering in \S\ \ref{sec:clust}.

\subsection{Simulated CMDs}
\label{sec:sim_CMD}
Our primary method to constrain the IMF is to compare the observed MS luminosity function to simulated MS luminosity functions. In this analysis we use the $g$-band MS luminosity function from the $(g-V)$ CMD for two reasons. First, the $(g-V)$ CMD has a MS with a more distinct separation from the rest of the CMD, including blue-loop stars, compared to the $(V-I)$ CMD. Second, in order to get a comparable selection in the $(V-I)$ CMD, avoiding RGB star contamination, we cannot go as far down the MS as we can in the $(g-V)$ CMD. This results in half the MS stars seen in the $(g-V)$ CMD ($\rm N_{obs}=181$). The $(g-I)$ CMD has a wider colour baseline than the $(g-V)$ CMD but does not improve how far down the MS we observe, resulting in essentially the same number of MS stars in the selection box.

To simulate the CMDs we begin with an assumed IMF and adopt a constant SFR (as justified in \S\ \ref{sec:SFH}) over 200 Myr, and use these to randomly  assign an initial mass and formation time to each star. For each set of input IMF parameters we produce an ensemble of 100 simulations, each with $5\times10^5$ stars, to produce good average test statistics. Our simulations assume that the stars are: single and non-rotating;  have uniform metallicity and dust extinction  \citep[$\rm Z=0.4\, Z_{\sun}=0.008$ and $E(B-V)=0.45$ mag for the outer-disc region,][]{Werk:2010fn}; and form randomly but continuously (not clustered in space or time). Except for the constant SFH, these are common assumptions of CMD analyses in nearby galaxies \citep[e.g.][]{Williams:2008ic, Williams:2010hw, Weisz:2011hn, Lianou:2013ej, Annibali:2013ie, Meschin:2014bd}. In $\S$ \ref{sec:caveats} we discuss the biases these assumptions may induce in our results.

We use the Geneva stellar evolutionary libraries since they cover our desired stellar-mass range ($1<\rm M/M_{\sun}<120$) at our assumed metallicity \citep[Z=0.008,][]{Schaerer:1993wj}. We interpolate between the available mass tracks to evolve the stars to an observation time $t_{obs}=200$ Myr, which is greater than the MS lifetime, $\sim$160 Myr of the lowest mass stars in our MS selection box. The extended simulation time allows for contaminations by older stars nominally outside the selection box, but scattered in by photometric errors. These contaminations should be matched by the systematic effects of the observed photometric errors. We then convert the evolutionary tracks to the ACS VEGAMAG system and apply foreground dust using the transformations of \citet{Sirianni:2005it}. We model photometric errors by adding magnitude perturbations to the model photometry, which have a Gaussian distribution with a width set by the photometric errors from the artificial star tests. We randomly remove stars according to the completeness factors derived from the artificial star tests. 

The resulting simulated $(g-V)$ CMD is comprised of all stars that are `alive'  at the end of each simulation \citep[that is they have an age since birth less than the lifetime given by][]{Schaerer:1993wj}. The simulated stars that are contained in the MS selection box shown in Fig. \ref{fig:CMD} are included in our statistical analysis. This selection is done regardless of whether the true age of the star is less than the MS lifetime of the star. Thus, in principle, the simulations suffer the same amount of contamination by non-MS stars as the observations. The distribution in the $g$-band luminosity of the stars within this selection box is taken to be the MS luminosity function, which we then compare to the observed MS luminosity function. Of the $5\times10^5$ stars in each simulation the number measured in the MS selection box depends on the IMF parameters. For a constant SFR this ranges from $\sim700$ ($\alpha = -3.95, \,\rm M_u=15\, M_{\sun}$) to $\sim6700$ ($\alpha  =-1.95, \,\rm M_u=120 \,M_{\sun}$). For consistency we randomly select 500 stars from within the MS selection box to compare to observations.  Our set of simulated CMDs have upper-mass limits that correspond to the high-mass Geneva evolutionary models \citep[i.e. 15, 20, 25, 40, 60, 85, 120 $\rm M_{\sun}$, ][]{Schaerer:1993wj} and IMF slopes ranging from -1.95 to -3.95 in steps of 0.1.

\subsection{Statistical test}
\label{sec:stat}
We compare the observed and simulated MS luminosity functions to determine if the observed data could have come from the the simulated MS luminosity function. 
We adopt the two sample Kolmogorov--Smirnov (K--S) test as our measure of how well the observed data match simulations. We use this approach because it does not require binning the data, or adopting a functional form for the distribution as would $\chi ^2$ minimisation. We also experimented with using the Kuiper test, which in principle should be more sensitive to the extremes of the distribution than the K--S test. However, we found this test prefers results weighted towards the low luminosity end of the distribution, typically over predicting the number of high-luminosity stars compared to observations.

To determine which models fit these data best we minimize the K--S test statistic D, which is the maximum deviation between the two distributions. Smaller values of D indicate that the two distributions are similar. We produce an ensemble of 100 simulations for each set of input IMF parameters to reduce stochastic effects. We take the mean D values at each IMF grid point of $\alpha$ and $\rm M_u$ as the value for that point. The grid point with the lowest mean D values represents the best-fitting IMF parameters.

\subsection{IMF constraints from the MS luminosity function}
\label{sec:IMF_constraints}
\begin{figure}
\centering
% contour_plot.pro
\includegraphics[width=85mm, height=50mm]{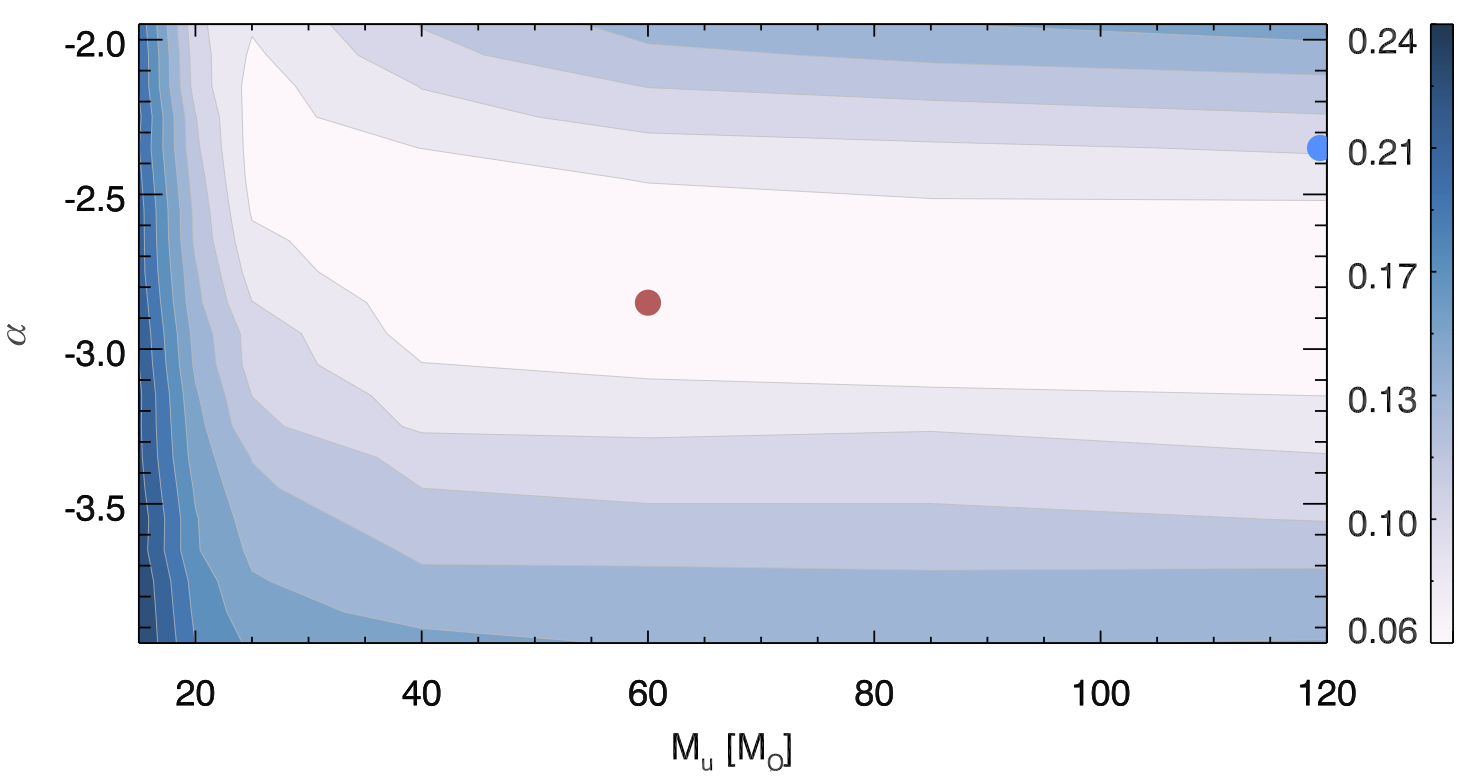}
\caption{Contour plot showing the mean test statistic D for each set of IMF parameters as a function of IMF slope ($\alpha$) and upper-mass limit (M$_{u}$). The red point shows the best-fitting IMF parameters and the blue point shows the location of a standard Kroupa IMF ($\alpha=-2.35$, M$_{u}=120$ M$_{\sun}$) for comparison. The bar to the right colour-codes the value of the test statistic D.}
\label{fig:contour_const}
\end{figure}

Fig. \ref{fig:contour_const} shows the results of the statistical tests as a contour plot of the statistic D in the plane of the free parameters, the IMF slope ($\alpha$), and upper-mass limit (M$_{u}$). The best-fitting parameters are $\alpha=-2.85 $ and $\rm M_{u}=60\,M_{\sun}$ (hereafter the best-fitting IMF). The contour plot also reveals a local minimum near $\rm M_{u}=25\,M_{\sun}$ and $\alpha=-2.35$ and extends to encompass $40 <\rm M_{u}/M_{\sun}<120$, indicating that the upper-mass limit is not well constrained. 

\begin{figure}
\centering
% compare_plot_all.pro
\includegraphics[width=85mm, height=100mm]{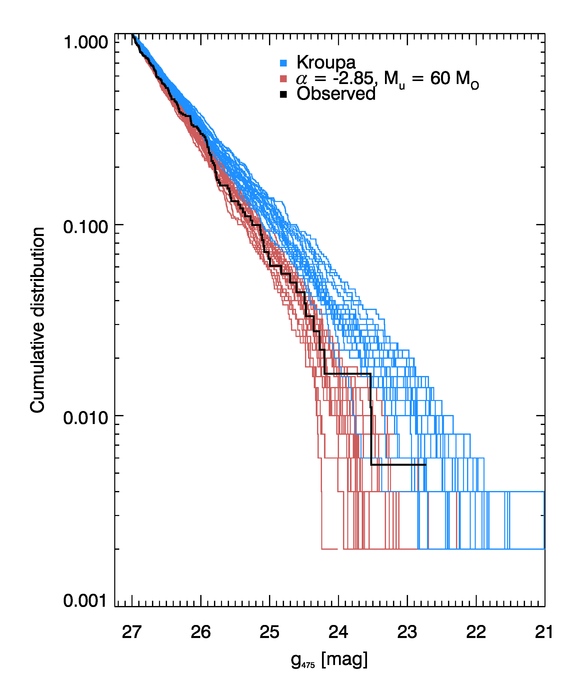}
\caption{Comparison between the observed MS luminosity function (black) and 20 best-matching realisations for a Kroupa IMF (blue) and our best-fitting IMF (red). This plot shows the normalized cumulative distribution for each MS luminosity function.} 
\label{fig:best_const}
\end{figure}

Fig. \ref{fig:best_const} compares the observed and simulated MS luminosity functions, plotted as normalized cumulative distributions, for the 20 best-matching realisation with $\alpha=-2.85$ and $\rm M_u=60 \, M_{\sun}$, along with the 20 best-matching Kroupa IMF realisations. By eye one can clearly see that the best-fitting IMF is preferable to the Kroupa IMF. We list the K--S test statistics and p-values for the best-matching realisations  for the Kroupa and our preferred IMF in Table \ref{tab:stats}.

\begin{table}
\centering
\caption{Results from the K--S test for the best matching realisation for $\alpha=-2.85$ and $\rm M_u=60\, M_{\sun}$, and the best matching Kroupa realisation compared to the observed MS luminosity function. Where p is the p-value, and D is the maximum deviation between the observed MS luminosity function and the simulated MS luminosity function for the K--S test.}
\label{tab:stats}
\begin{tabular}{|c|c|c|}
\hline
IMF & p & D \\ 
($\alpha$, $\rm M_u$) & &  \\ \hline \hline
-2.85, $\rm60\,M_{\sun}$ & 0.96 & 4.3$\times10^{-2}$ \\
-2.35, $\rm120\,M_{\sun}$ & 0.32 & 8.1$\times10^{-2}$\\ \hline
\end{tabular}
\end{table}

We ran simulations to determine how well this technique recovers known IMF parameters, and to estimate the uncertainties in these parameters. To do this a simulated stellar population (as described in $\S$ \ref{sec:sim_CMD}) is used as the `observed' data (with known $\alpha$ and $\rm M_u$). We randomly select MS stars, equal in number to those observed, from the simulation to produce an `observed' MS luminosity function. We use the same procedure as described in \S\ \ref{sec:stat} to determine the best-fitting parameters of the `observed' MS luminosity function. We then repeat this for the other 99 simulations with the same $\alpha$ and $\rm M_u$ as the `observed' MS luminosity function. The variance of the best-fitting values is used as the uncertainty for the IMF parameters.

For an input Kroupa IMF the average recovered IMF parameters are $\alpha= -2.28\pm0.28$ and  $\rm M_{u}=78 \pm 31 \, M_{\sun}$. For our best-fitting IMF parameter, we determined the average recovered IMF slope via the K--S test to be $\alpha = -2.75 \pm 0.36 \,$ and the upper-mass limit to be $\rm M_{u} = 70 \pm 36,M_{\sun}$. In these tests the recovered $\alpha$ is within 1$\sigma$ of the input value, although consistently less steep than the input. When the input $\rm M_u$ is 120 $\rm M_{\sun}$, the recovered value is less than this, because this is the maximum allowed value, otherwise it is within 1$\sigma$ of the input. The large 1$\sigma$ uncertainty for $\rm M_u$  is consistent with the elongated region seen in Fig. \ref{fig:contour_const}. By varying the input upper-mass limit, we determine that we can constrain $\rm M_u$ but only if $\rm M_u<25\,M_{\sun}$. Tighter constraints at higher $\rm M_u$ are not possible using the $g$ versus $g-V$ CMD alone.

Nominally the p-values in Table \ref{tab:stats}, do not rule out a standard Kroupa IMF, so we consider whether the simulations provide insight into the applicability of this standard IMF. We determine the distribution of D statistics for a Kroupa IMF by comparing the MS luminosity functions of all the Kroupa simulations ($10^4$ tests). We then compare the observed MS luminosity function of NGC 2915 to the Kroupa IMF simulations to produce a comparable observation versus simulation D distribution. Since there is only one observation set this allows 100 tests. If the observations match the simulations we should find that 68\% of the observed-simulation tests should have D values less than the 68th percentile in the simulation-simulation tests, with the 68th percentile used as the standard marker for a 1$\sigma$ match. We find that none of the D values are lower than the 68th percentile value. This implies that our observations exclude the Kroupa IMF simulations to a probability $\la 1$\%.

Using the method outlined in Appendix A and our best-fitting IMF parameters we determine the SFR integrated over the observed field to be $8.1 \times 10^{-3}\, \rm M_{\sun}\, yr^{-1}$ and \SFI\ of $2.9 \times 10^{-4}\, \rm M_{\sun}\, yr^{-1}\, kpc^{-2}$. Here we use our best-fitting model from the MS luminosity function analysis for the upper IMF ($m \geq 1\,\rm M_{\sun}$), and a standard Kroupa IMF for $\rm 0.08\, M_{\sun}<m<1\, M_{\sun}$. In comparison using a standard Kroupa IMF for all masses would result in SFR and \SFI\ being a factor of 3.2 lower. The \SFI\ value is corrected to be face-on using the inclination listed in Table \ref{tab:info}.

\subsection{Efficacy of the MS luminosity function analysis}
We find that an optical MS luminosity function analysis alone does not provide a good constraint to both $\alpha$ and $\rm M_u$. This is a well known problem: you cannot determine the mass or spectral type of a star based on broadband optical photometry alone \citep{Massey:1998tx, Massey:2013fv}. Since high-mass MS stars emit most of their energy in the UV there is not much change in their optical colours during their MS lifetimes. There is however, significant change in their optical luminosity, which causes the evolutionary tracks to be steep (see Fig. \ref{fig:zoom_HII}). This problem is exemplified by looking at the difference in the optical properties of a newly formed $\rm 65\,M_{\sun}$ MS star and a 6.3 Myr old $\rm 25\,M_{\sun}$ MS star. The $\rm 65\,M_{\sun}$ star has an absolute \textit{V}-band magnitude of $M_V=-5.28$, while the $\rm 25\,M_{\sun}$ has an absolute \textit{V}-band magnitude of $-5.19$ \citep{Schaerer:1993wj}. The small difference in absolute \textit{V}-band magnitude between tracks of vastly different stellar mass and age stars makes it hard to optically discriminate between stellar masses. Optical spectroscopy is required for precise determination of spectral type, and thus stellar masses \citep{Massey:1998tx, Massey:2013fv}. Alternatively, fitting of the UV to optical spectral energy distribution of stars provides an efficient means to improve upon optical photometry only \citep{Bianchi:2006iq, Bianchi:2012he}. The additional data required for this analysis has not been obtained for NGC 2915.

The brightest star in the observed CMD is separated by $\sim$0.8 mag from the other MS stars (see Fig. \ref{fig:HII_reg}). As shown in Fig. \ref{fig:massive_5} and noted in \S\ \ref{sec:HII} it appears to have a few faint stars projected near it. So perhaps this is not a star but the centre of a compact cluster. Even if it is a star in NGC 2915, in a few Myr it may disappear in a supernova explosion, so the existence of such a bright source may be fortuitous. Therefore, it is worthwhile considering how sensitive our results are to this one source. We removed the most luminous star from the observed MS luminosity function to determine the dependence of our results on the brightest star. We determine the best-fitting IMF parameters for the CMD only analysis using the new MS luminosity function with this source removed. We find that the removal of the brightest star has no effect on the recovered IMF parameters.

\begin{figure}
\centering
\includegraphics[width=85mm, height=70mm]{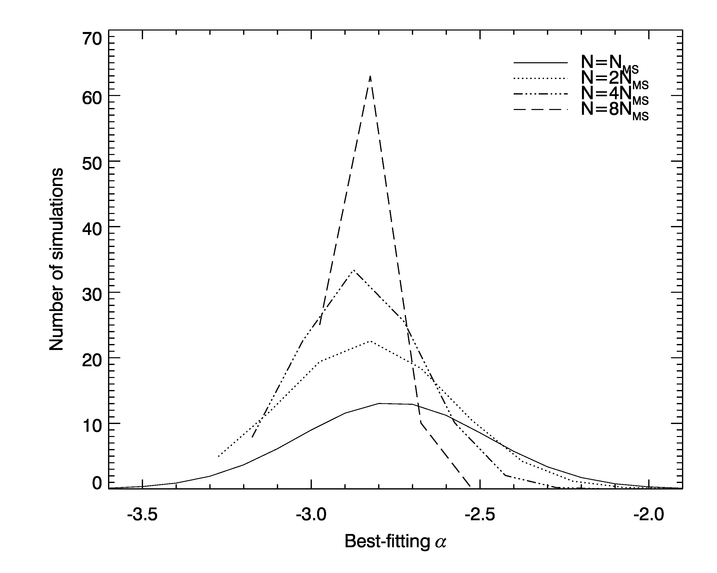}
% CMD_uncert_gaus
\caption{The effect of increasing the number of MS stars on the recovered IMF slope for an input IMF slope of $\alpha=-2.85$. The number of MS stars is varied from $\rm N_{MS}=N_{obs}=181, $ (dash) to 8 times this value (solid). The y-axis shows the number of simulations that have the best-fitting $\alpha$ given by the x-axis. A Gaussian fit to these results becomes narrower, indicating a reduction in the uncertainty of the recovered values.}
\label{fig:val_NMS}
\end{figure}

\subsection{Uncertainty of IMF parameters}
We ran simulations to determine the effect of increasing the number of observed MS stars, on the derived IMF parameters from the MS luminosity function analysis. We randomly select the required number of `observed' MS stars ($\rm N_{ MS}$) from the simulated MS stars in multiples of the number of observed MS stars in our data. We then determine the best-fitting IMF parameters using the method described in \S \ref{sec:stat}. Fig. \ref{fig:val_NMS} shows Gaussian fits to the histogram of recovered $\alpha$ values as a function of $\rm N_{ MS}$. The resultant fitted Gaussian dispersions are 0.28, 0.25, 0.18, 0.09 for 1,2,4, and 8 times the number of observed MS stars respectively, hence the uncertainty in $\alpha$ decreases as $\sim \rm N_{MS}^{-1/2}$ over this range. However, changing $\rm N_{ MS}$ does not strongly affect the uncertainty of $\rm M_u$; over the whole range of  $\rm N_{MS}$ trialled the $1 \sigma$ uncertainty in $\rm M_u$ reduces from  34 to 22 $\rm M_{\sun}$. The above results allow us to estimate the best achievable accuracy for $\alpha$ allowed by a pure CMD based analysis of the MS luminosity function, given the number of MS stars. By applying the SFL results (\S\ref{sec:SFlaw} below) to the observed \HI\ distribution of \citet{Elson:2010iv} we estimate the total number of MS stars for radii 45 -- 257 arcsec to be $\rm N_{MS} \sim 870$, which would allow the best possible uncertainty in $\alpha$ of 0.18 if the whole outer disc of NGC 2915 were to be observed with the \textit{HST} in the manner done here.

If we apply a similar CMD only analysis to other galaxies at a similar distance with a single fixed field of view detector, like ACS/WFC, then the limiting accuracy in $\alpha$ depends on the average $\Sigma_{SFR}$. For galaxies at a similar distance to NGC 2915, larger number of observed MS stars could be achieved by observing a larger area, looking at higher surface brightness regions, looking at galaxies more favourably inclined, or by stacking results of different galaxies. Since the IMF may vary as a function of galaxy surface brightness \citep{Meurer:2009gp} caution needs to be applied when stacking results from multiple galaxies covering a range of \SFI. The uncertainty on $\alpha$ likely depends not only on the number of observed MS stars but on the depth of the observed data, filters used, and other systematic effects like crowding. On the other hand, increasing the number of observed MS stars does not improve the uncertainty of $\rm M_u$. Thus, it seems unlikely that any similar MS luminosity function analysis can improve the degeneracy we find in $\rm M_u$ if only optical filters are used. Constraining both $\rm M_u$ and $\alpha$ requires other techniques, such as \Ha\ constraints (as explored next).

\subsection{IMF constraints from \Ha\ flux}
\label{sec:CMD_Ha}
Here we use the \Ha\ flux of \HII-1 from \citet{Werk:2010fn} to provide an independent constraint on the upper end of the IMF. If this is a normal \HII\ region with case-B recombination then the \Ha\ flux measures the total ionizing flux and thus the O star content of \HII-1. We normalize the measured \Ha\ flux by the sum of the STMAG \textit{V}-band flux densities of the stars in our MS selection box, yielding a `pseudo' \Ha\ equivalent width of $w_{\rm{H} \alpha} = 28.1 \pm 4.8$ \AA. This is a pseudo equivalent width because the flux density is not measured at the same wavelength as the \Ha\ emission, and the CMD selection box misses the fainter stars associated with star formation as well as older stellar populations. Since the spectra of hot young stars are fairly constant over the optical part of the spectrum, incompleteness is the major effect, hence $w_{\rm{H} \alpha}$ over estimates the true equivalent width.

The accuracy of the match in $w_{\rm H\alpha}$ is an important parameter in determining the constraints. The constraint is too tight and arbitrary if we require the match to agree with the observational uncertainty; with better observations presumably we could arbitrarily tighten the error in $w_{\rm H\alpha}$, and it would become increasingly hard to match the narrow window in this quantity. Therefore, we extend the uncertainties to range from 0.5 $w_{\rm H\alpha}$ -- 2 $w_{\rm H\alpha}$, and adopt  $14 \rm{\AA} <$ $w_{\rm{H} \alpha}<56 \rm{\AA} $ as the constraints we require for a match to the \Ha\ observations. 

\begin{figure}
\centering
\includegraphics[width=85mm, height=50mm]{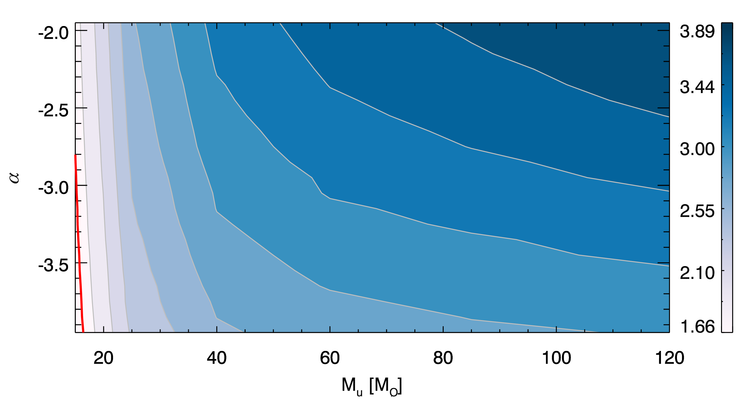}
% Conti et al EW contour plot in evernote
\caption{Contour plot showing the log of the average pseudo \Ha\ equivalent width of the simulations as a function IMF parameters. Only IMFs with a $\rm M_u<15\, M_{\sun}$ and $\alpha< -2.85$ have an average $w_{\rm H\alpha}$ within the limits of the observed $w_{\rm H\alpha}$ from \HII-1 ($1.1 < \log(w_{\rm{H} \alpha}/\AA )<1.7 $, shown as the red contour).}
\label{fig:EW_cont}
\end{figure}

For each of our previously produced simulations we randomly select stars that are within our selection box, to match the observed number of stars, and model the $w_{\rm H\alpha}$. We re-select the same number of MS stars from each simulation 20 times to improve the statistics (i.e. $n \sim 2000$ samples per set of input IMF parameters). We calculate the ionizing photon rate of the simulated MS stars using their initial masses and interpolating data for dwarf stars in the mass range $\rm 7.5\, M_{\sun}-80\, M_{\sun}$ and extrapolating for masses $>80\, M_{\sun}$ from Table 3.1 in \citet{Conti:2008ur}. We then convert this to \Ha\ flux and obtain the $w_{\rm H\alpha}$ for each simulation by summing the \Ha\ and $V$-band fluxes over all MS stars. Using this method we find that only simulations with an upper-mass limit $\rm M_u<20\, M_{\sun}$ match the constraints applied to $w_{\rm H\alpha}$, as shown in Fig. \ref{fig:EW_cont}, implying a mismatch between the CMD and \Ha\ observations. The \Ha\ observations imply there is a single O star in the field, while the CMD implies there are many stars with masses high enough to produce significant \Ha\ emission (see Fig. \ref{fig:CMD}).

Before discussing the dichotomy in the results, we note that the slit spectra from the Magellan Baade telescope by \citet{Werk:2010fn} were re-analysed to determine if they could constrain the amount of \Ha\ diffuse ionized gas (DIG) in our field. A total of 13 slitlets each 10 arcsec long and 0.7 arcsec wide project on to the \HI\ disc of NGC 2915 but avoid readily apparent sources, at least over half their lengths. This includes six slitlets within the ACS field. Two slitlets placed outside the \HI\ extent of NGC 2915 were used for sky subtraction. We do not detect \Ha\ emission to a 3$\sigma$ detection limit of $S_{\rm H\alpha} \leq 3.5 \times 10^{-18}\, \rm{erg\, cm^{-2}\, s^{-1}\, arcsec^{-2}}$. This is derived from a weighted average of all measurements within the ACS aperture. All individual measurements have fluxes less than the 3$\sigma$ limit for that particular slit whether or not they are within the ACS aperture, with some fluxes nominally negative. As noted by \citet{Werk:2010fn} the main limitation is systematic, due to using non-local sky spectra. How does this compare to what we expect? Taking the estimated \SFI\ from the CMD only analysis assuming a Kroupa IMF (Table \ref{tab:ratios} Appendix) and applying the appropriate inclination, dust absorption, and the \Ha\ luminosity conversion of \citet[][corrected to a Kroupa IMF]{Kennicutt:1998id} we estimate we should observe $S_{\rm H\alpha} \sim 2.6  \times 10^{-18}\, \rm{erg\, cm^{-2}\, s^{-1}\, arcsec^{-2}}$ from the observed blue stars assuming the emission is spread evenly over the field. Hence, the existing spectra are consistent with no DIG emission, but represent only a $\sim 3\sigma$ non-detection. We use the $3\sigma$ detection limit to place an upper limit of $w_{\rm H\alpha}<8771\,\rm{\AA}$ on the \Ha\ DIG. Thus the Magellan spectra do not provide a significant constraint on the upper-end of the IMF.

\subsection{Dichotomy between observed \Ha\ emission and CMD}
\label{sec:clust}
If we assume that all of the \Ha\ emission is contained in \HII-1 then there is a mismatch between the CMD and \Ha\ observations. The observed CMD in Fig. \ref{fig:CMD} implies multiple high-mass ($\rm M_\star > 15\, M_{\sun}$), MS stars that are capable of ionizing the ISM. Yet, the observed \Ha\ emission from \HII-1 requires a single late O star with $\rm M_{star} \sim 19\, M_{\sun}$, or more ionizing stars of lower masses. Plausible explanations for this dichotomy include: (1) stochasticity due to the small number of high mass stars expected; (2) many of the objects we identify as upper-MS stars are actually multiple star systems; (3) the ionizing flux from this field is not all captured by \HII-1 with some of the ionizing radiation escaping  to form undetected DIG (or escaping the galaxy entirely). Of course it is also possible that some combination of these scenarios is the correct explanation for this dichotomy. We now consider each of these explanations in turn.

\textbf{Stochasticity.}
Due to stochastic sampling of the IMF and the decreasing MS lifetime with increasing stellar mass, the probability of observing O stars reduces with SFR. If the SFR is sufficiently low the probability of observing an O star and hence \Ha\ emissions becomes very low, despite ongoing star formation. However, the problem is that for either a Kroupa IMF or the IMF parameters of our best-fitting MS luminosity function we expect more O stars to be present than implied by the \HII-1 results. For our best-fitting IMF (or Kroupa) our simulations show that on average $\sim$26\% (34\%) of the stars in our MS selection box should be real MS stars with $\rm M_{\star}>15\,M_{\sun}$, hence capable of ionizing the ISM. Scaled to the 181 stars observed in the MS selection box we expect 48 (63) to be ionizing; and the Poissonian fractional uncertainty to be 0.14 (0.12), while the probability to have just one O star present is $6.8\times 10^{-20}$ ($2.7\times 10^{-26}$). One way to improve these long odds is if star formation is clustered in time, that is through a series of mini-bursts, each one representing the formation of one stellar cluster, group or association. Note this case is separate from the formation of tight bound clusters, which we consider below. \citet{Thilker:2007ho} produced simple Monte--Carlo simulations for continuously star-forming clustered populations to determine the number of observed O stars. The probability of finding just one ionizing cluster, e.g. like the association around 
HII-1, is $\sim$ 1\% (assuming a Salpeter IMF with a cluster mass function (CMF) slope of -2, allowing clusters up to $\rm 10^6\, M_{\sun}$). This is for a SFR of $\rm SFR\sim10^{-3}\, M_{\sun}\, yr^{-1}$, which is lower than our estimated SFR of $\rm 2.45\times10^{-3}\, M_{\sun}\, yr^{-1}$ (assuming a Kroupa IMF), meaning that the actual probability is lower. While still a low probability, it is sufficiently high that temporally clustered star formation becomes a plausible explanation or partial explanation.

\textbf{Unresolved clusters.}
If the brightest sources in the MS selection box are not single stars then they could hide a lot of stellar mass. For example with the standard cubic stellar luminosity--mass relation one would need a thousand non-ionizing stars with mass 6 $\rm M_{\sun}$ to account for the same luminosity as one 60 $\rm M_{\sun}$ star. In our data we are unable to resolve systems much smaller than the pixel scale of 0.99 pc. The entire binary separation distribution for Galactic stars fits into one tenth this size \citep{Marks:2014eb}, hence all binaries will appear single in our observations. Thus, while all the sources within our selection box appear to be single many may be binaries. Earlier (\S\ \ref{sec:HII}) we noted that the brightest source in Fig. \ref{fig:massive_5} may be part of a compact group. Following the work of \citet{Hill:2006et} we expect other low mass clusters to have a similar size. They find an average core radius of young LMC clusters to be 4.2 pc. Since the core radius is similar to a half light radius then we expect young clusters to have diameters of 8 pixels, readily detectable by the WFC. But some of their LMC clusters have core radii as low as 0.61 pc, which would be hard to distinguish from a single star. So it seems reasonable to expect that at least some of the high-luminosity sources in the MS selection box are multiple star systems. Since we underestimate the mass by assuming each source is single (as noted above) then the single star assumption likewise overestimates the importance of high-mass stars. This has also been noted by \citet{Grillmair:1998fl}, \citet{Sagar:1991ub} and, \citet{Lamb:2012tc}. Thus, the resultant IMF will be even more deficient in high-mass stars than derived in \S\ \ref{sec:IMF_constraints}. Almost all resolved stellar population studies directed at studying the IMF at distances similar to NGC 2915 also assume the stars are single \citep[e.g.][]{Tosi:1991ed,Greggio:1993dl,Marconi:1995jp,Crone:2002bi,Annibali:2003jh}; hence those studies may also be overestimating the content of the highest mass stars in galaxies.

\textbf{Diffuse Ionized Gas (DIG).}
If the \HII\ regions are `leaky', ionizing photons could escape and not be counted, this would break the assumption of case B recombinations (all ionizing photons are absorbed locally). The photons that leak out could be absorbed further afield in NGC 2915 to be part of the DIG, or perhaps escape the galaxy totally. DIG emission have been found to comprise a large fraction of the \Ha\ flux of galaxies; for example \citet{Oey:2007cj} determined that $\sim60$\% of \Ha\ emission from local \HI\ selected galaxies is DIG.  If there is significant DIG or the ionizing radiation is escaping the galaxy entirely then the limits imposed on $w_{\rm H\alpha}$ by the Magellan spectra allow all IMF models we have trialled. However, they do not negate the MS luminosity function fit, which shows that an IMF with $\alpha=-2.85$ and $\rm M_u=60\, M_{\sun}$ is preferred i.e. modestly deficient in the highest mass stars. We note that a leaky \HII\ region explanation has also been posited to explain the case of low H $\alpha$/FUV in low luminosity LSB galaxies by \citet{Hunter:2010fx}.  If high-mass stars form outside of \HII\ regions or the regions are leaky then star formation studies limited to just \HII\ regions will undercount the total SFR \citep[e.g.][]{Kennicutt:1983cq, Calzetti:2007en}. One major problem with the DIG explanation is that low luminosity \HII\ regions such as \HII-1 should not be leaky. \citet{Pellegrini:2012ii} show that \HII\ regions in the Magellan Clouds below $\log(\rm L$/erg s$^{-1})=37.0$ are optically thick and do not leak ionizing photons. Indeed \citet{Oey:2007cj} find that the leakiest galaxies have high surface brightness, not low surface brightness. This suggests that DIG should not dominate in the low surface brightness outer region of NGC 2915.  We also note that while individual stars have been detected outside of the core of \HII\ regions the Galaxy and the LMC \citep{deWit:2004fg, deWit:2005kd, Bressert:2012gt}, they make up only a small fraction of the total O star content \citep[i.e. about 4\% in the Galaxy according to][]{deWit:2005kd}.

\subsection{Caveats and limitations}
\label{sec:caveats}
Our results are dependent on the assumptions of our models and the corrections applied to our data. Here we review those assumptions and the effects on our results.
 
We model a uniform total extinction of $E(B-V)=0.45$ mag, determined by \citet{Werk:2010fn} for H {\sc ii}-1. Using the foreground extinction of $E(B-V)_{f}=0.275$ mag taken from \citet{Schlegel:1998fw} the average internal extinction for NGC 2915 is $E(B-V)_{i}=0.175$ mag. However, the ISM is particularly dense and clumpy where new stars are being formed. This is likely to add intrinsic scatter to the photometry, which we have not modelled. 

The evolutionary tracks we employ do not include stellar rotation, which extends the MS lifetime, induces stronger mass loss, and causes higher effective temperatures and luminosities compared to non-rotating stars of the same initial masses \citep{Levesque:2012bv}. According to \citet{Leitherer:2014tv} the differences between rotating and non-rotation stellar populations are more pronounced in later evolutionary phases and in solar metallicity environments, compared to sub-solar metallicity environments. Thus, neglecting stellar rotation is likely to bias the results to have a higher upper-mass limit or shallower slope than actually exists, although the effect may be small due to the low metallicity of NGC 2915.

As noted in \S\ \ref{sec:clust} we neglect binaries and higher order multiples in our analysis and this will result in overestimating the content of the highest mass stars. Hence our results will bias the IMF slope to shallow values and the upper-mass limit to higher values compared to reality. Likewise in \S\ \ref{sec:IMF_constraints} we show using simulations that our numerical method tends to recover IMF slopes that are shallower than the simulations.

In summary, neglecting stellar rotation and multiple stars systems will both bias the results to appear to have more of the highest mass stars than actually exist, while our method of analysis also results in IMF slopes that are too shallow. Since the best-fitting MS luminosity function results is an IMF that is already deficient in high-mass stars compared to the Kroupa IMF, the actual deficiency in high-mass stars is likely to be more extreme than our best-fitting IMF. Unless there is significant but undetected DIG content the \Ha\ analysis indicates that the IMF is even more deficient still in stars with masses $\rm >25\, M_{\sun}$.

\begin{figure}
\centering
% HI_blue.pro
\includegraphics[width=90mm, height=90mm]{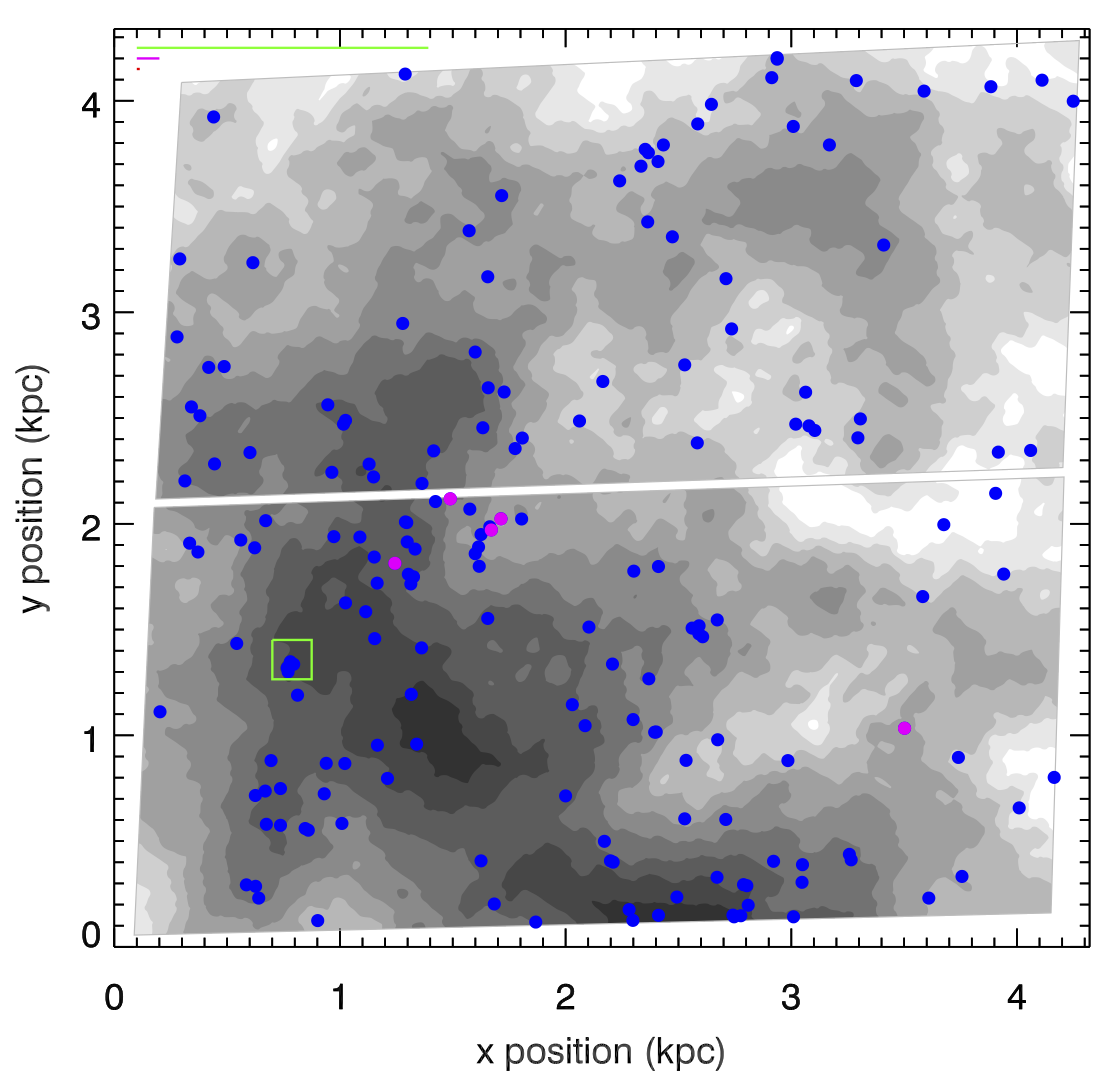}
\caption{Comparison of the distribution of young stars with the \HI\ gas distribution. The MS stars, as identified in the $g-V$ CMD, are shown in blue, the five most luminous MS stars are shown in magenta, and the \HII-1 stars are located within the green box. The inclination corrected \HI\ radio contours, from the natural weighted \HI\ total intensity map of \citet{Elson:2010iv}, are plotted for the WFC regions. The  \HI\ data contour levels are 1.5, 2.3, 2.7, 3.2, 3.8, 4.5, 5.2, 6.2, 7.2, 8.5, 10.2 $\rm M_{\sun}\, pc^{-2}$, with the darkest grey-scale colour corresponding to the densest region. The MS stars have a clumpy distribution that appears to follow the \HI\ gas distribution. The green, magenta and orange vectors, shown in the upper left corner, indicate the maximum, average and minimum distance a MS star could travel from where it originally formed during its lifetime respectively.}
\label{fig:HI_dist}
\end{figure}

\section{Star formation law}
\label{sec:SFlaw}
\subsection{Local star formation law}

\begin{figure*}
\centering
\includegraphics[width=85mm, height=85mm]{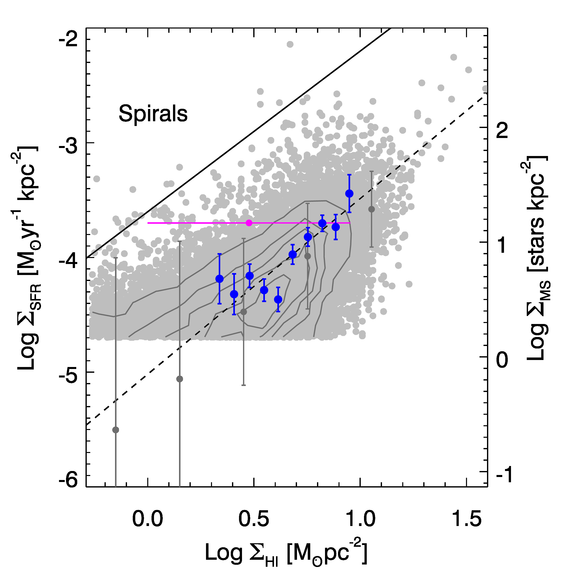}
\includegraphics[width=85mm, height=85mm]{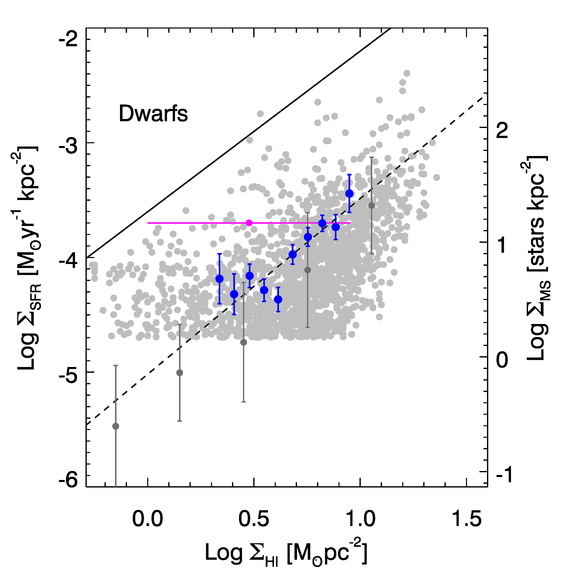}
% MSden_to_SFRden.pro
\caption{The star formation law for the outer disc of NGC 2915 (blue points) compared to the outskirts of spiral (left) and dwarf galaxies (right) shown as grey points \citep[from][]{Bigiel:2010jc}. All data have been corrected to face-on quantities. \SFI\ for NGC 2915 was determined from MS star density (right axis), with errors determined as Poisson counting errors.  A linear ordinary least-squares fit to our data is plotted as the dashed line. This fit yields a slope of $\rm N=1.53 \pm 0.21$ and $\rm A=(9.6 \pm 3.3)\times 10^{-6} \, M_{\sun}\, yr^{-1}\, kpc^{-2}$. The dark grey contours in the left panel enclose where 16\%, 35\%, 53\%, 67\% and 86\% of data are located. The dark grey data points with error bars, in both panels, represent median values and the scatter from \citet{Bigiel:2010jc}. NGC 2915 displays properties of both spiral and dwarf galaxies, due to its differing optical and \HI\ properties. The  \SFI\ of NGC 2915 is slightly above the median relationships for the outskirts of spiral and dwarf galaxies but well within the scatter of both datasets. We have included the global Schmidt law for star-forming galaxies \citep{Kennicutt:1998id} for reference (solid line). The \SFI\ determined from \textit{GALEX} UV data by \citet{Werk:2010fn} for 44--150 arcsec is shown in magenta (corrected to a Kroupa IMF), with error bars that indicate the range of radially averaged \HI\ column densities the \SFI\ is derived over.}
\label{fig:HI_SFR}
\end{figure*}

The position of the MS stars with respect to the \HI\, provides a means to examine the nature of the SFL in the outer disc of NGC 2915. Fig. \ref{fig:HI_dist} shows the distribution of MS stars and the \HI\ data from \citet{Elson:2010iv}. The MS stars have a clumpy distribution that follows the \HI\ gas distribution, indicating recent star formation and \HI\ gas may be correlated. The correlation between MS star surface density (right axis), and the \HI\ surface density for the outer-disc region of NGC 2915 is shown in Fig. \ref{fig:HI_SFR}. We calculate the inclination corrected \HI\ surface brightness at the position of each star using the natural weighted \HI\ total intensity map from \citet{Elson:2010iv}, which has a resolution of 17.6 arcsec  ($\sim350$ pc). We determine the surface density of MS stars ($\Sigma_{\rm MS}$) by counting the number of MS stars in 0.1 dex bins of $\Sigma_{\rm HI}$, and dividing by the area in our ACS image that have the same $\Sigma_{\rm HI}$. Thus our method is equivalent to that in the early study of \citet{Sanduleak:1969hn}. The stellar density is converted to a star formation intensity (left axis) assuming a Kroupa IMF following Appendix A. We use a Kroupa IMF here to allow better comparison with previously published results. Our data are shown in blue along with a linear ordinary least-squares fit (dashed line). This fit yields a slope of $\rm N=1.53 \pm 0.21$ and $\rm A=(9.6 \pm 3.3)\times 10^{-6} \, \rm M_{\sun}\, yr^{-1}\, kpc^{-2}$.  If our best-fitting IMF is adopted, then A is a factor of 4.5 higher.

Fig. \ref{fig:HI_SFR} compares our results to the `standard' integrated SFL of \citet{Kennicutt:1998id} as well as the results on low-density star formation in the outer discs of dwarf and spiral galaxies by \citet{Bigiel:2010jc}. This plot demonstrates that the outer disc of NGC 2915 has a star formation intensity significantly depressed compared to the extrapolated standard SFL, but slightly higher compared to what is observed in the outer discs of other spiral and dwarf galaxies. The comparison to both spirals and dwarfs is relevant since NGC 2915 exhibits properties of both due to its differing optical and \HI\ properties \citep{Meurer:1994kv, Meurer:1996ca}. We note that data displayed by us and \citet{Bigiel:2010jc} differs: \citet{Bigiel:2010jc} shows star formation intensity derived from UV surface brightness and \HI\ surface surface mass density averaged over square boxes of size 750 pc, whereas we derive our correlation from MS star counts compared to the local \HI\ surface mass density.

This method assumes the the stars are currently located in the gas cloud in which they formed. Even though the MS stars and \HI\ column density are correlated, there are relatively few MS stars located in the densest \HI\ contours of Fig. \ref{fig:HI_dist}. This suggests that the MS stars are now located in regions of lower density than where they originally formed. This has the effect of elevating the SFL above the median values of \citet{Bigiel:2010jc}, which can be seen in Fig. \ref{fig:HI_dist}. We place three vectors in Fig. \ref{fig:HI_dist} to show the maximum, average and minimum distance a MS star could travel from its birth location. We calculate the average distance by determining the average age of stars in out MS selection box from our simulations, assuming a Kroupa IMF, constant SFR and velocity equal to the \HI\ velocity dispersion. The minimum and maximum distances are based on the MS lifetime of a $\rm 120\, M_{\sun}$ and $\rm 4\, M_{\sun}$ star respectively.

\subsection{Fits to the star formation law}
Our results add to the relationship shown by \citet{Bigiel:2010jc} between \HI\ surface density and \SFI\ in the outer regions of galaxies. This differs from the SFL for the inner parts of galaxies where the \SFI\ correlates with \Htwo. This supports the idea that star formation in the outer regions of galaxies is regulated by the \HI\ column density \citep{Bigiel:2010jc}. \citet{Krumholz:2013kk} has modelled star formation intensity in regions where the ISM is comprised mostly of \HI, such as in dwarf galaxies and the outer discs of spirals, his results are in agreement with our findings and those of \citet{Bigiel:2010jc}. \citet{Krumholz:2013kk} postulates that the difference between the inner and outer SFLs is due to differences in the processes that regulate the conversion of  \HI\ to \Htwo, and thus the SFR \citep{Krumholz:2013kk}. The large scatter seen in the SFL of different galaxies could be explained by different concentrations of matter in the discs.

\begin{figure}
\centering
\includegraphics[width=88mm, height=75mm]{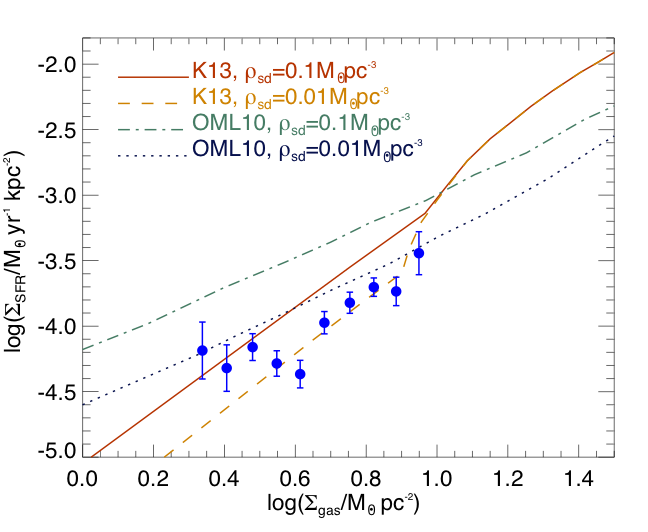}
\caption{Comparison of the observed SFL in the outer disc region of NGC 2915 with the star formation models of \citet[][K13]{Krumholz:2013kk} and \citet[][OML]{Ostriker:2010dm}. We assume that the total gas density is equal to the \HI\ gas density and a metallicity of $Z=0.4\, Z_{\sun}$. Our data is shown in blue and is the same as Fig. \ref{fig:HI_SFR}. For both model types we show results for adopted volume density of stars plus dark matter $\rm \rho_{sd}=0.1\,  M_{\sun}\,pc^{-3}$ and $\rm \rho_{sd}=0.01\,  M_{\sun}\,pc^{-3}$.} 
\label{fig:SF_law_C}
\end{figure}

We compare the observed SFL in the outer-disc region of NGC 2915 to the models of \citet[][hereafter OML]{Ostriker:2010dm} and \citet[][hereafter K13]{Krumholz:2013kk}, by assuming the total surface gas density (\HI\ + \Htwo) is equal to the surface density of \HI\ (i.e. \HI\  dominated). Fig. \ref{fig:SF_law_C} shows both models fitted with two different volume densities of stars plus dark matter $\rm \rho_{sd}=0.1\, M_{\sun}\, pc^{-3}$ and $\rm \rho_{sd}=0.01\, M_{\sun}\, pc^{-3}$ at a metallicity of $Z=0.4\, Z_{\sun}$. Our observations are consistent with the low density model of OML. While they slightly favour the low density model of K13 to the higher density option, the distinction is less clear for this model set. We conclude that the SFL fits imply a total mass density of stars and dark matter in the disc of $\rm \sim 0.01\, M_{\sun}\, pc^{-3}$, or perhaps a factor or few higher.

Are the density estimates from these models consistent with the IMF results? Limits on the projected stellar mass density due to the current SFR depend on the time-scale over which it occurs, i.e. from 1 orbit ($\sim 230$ Myr, minimum time-scale) to the Hubble time. This results in $\rm \Sigma_{\star,min}=4.2 \times 10^{-3} \, M_{\sun} \, pc^{-2}$ to $\rm \Sigma_{\star,max}=2.5 \times 10^{-1} \, M_{\sun} \, pc^{-2}$ if a Kroupa IMF is assumed; and $\rm \Sigma_{\star,min}=1.3 \times 10^{-2} \, M_{\sun} \, pc^{-2}$ to $\rm \Sigma_{\star,max}=8.1 \, M_{\sun} \, pc^{-1}$ for an IMF with $\alpha=-2.85$ and $\rm M_u=60\, M_{\sun}$. The corresponding volume density depends on the height of the disc. Often a typical scale height of $\sim100$ pc is assumed, adopting this and taking the total depth as twice that, we arrive at stellar mass volume densities: $\rm \rho_{\star,min}=4.2 \times 10^{-5} \, M_{\sun} \, pc^{-3}$ to $\rm \rho_{\star,max}=2.5\times 10^{-3} \, M_{\sun} \, pc^{-3}$ if a Kroupa IMF is assumed, and $\rm \rho_{\star,min}=1.3 \times 10^{-4} \, M_{\sun} \, pc^{-3}$ to $\rm \rho_{\star,max}=8.1 \times 10^{-3} \, M_{\sun} \, pc^{-3}$ if our preferred IMF is assumed. This latter $\rm \rho_{\star,max}$ is within 20\% of the low $\rm \rho_{sd}$ models that correspond best to our data, which we consider to be sufficiently close. We also estimate a crude maximum depth by positing that the observed \HI\ spiral arms \citep{Meurer:1996ca,Elson:2010iv} should not be taller (size normal to the disc) than they are wide (size in the plane of the disc). Then the observed arm width $\sim 500$ pc is the maximum depth and the above volume densities could be up to a factor of 0.4 times lower. Note that the CMDs in Fig. \ref{fig:CMD} indicate that the old stellar population should be significant, but we are not directly including it in these estimates; analysing these stars is outside the scope of this paper, and it is unclear whether they are in the disc or not. However, this population is implicitly counted in our mass estimates if it is in the disc, and star formation has been occurring for as long as a Hubble time. We see that if the star formation has been occurring over such a long time-scale at its present rate, the disc is thin, and our preferred IMF is employed, then there is enough baryonic matter in the disc to account for all of the required $\rho_{\rm sd}$ without requiring there to be disc dark matter to contribute to this density. 

\subsection{Total star formation in the outer disc}
We determine the total SFR integrated over the entire \HI\ disc of NGC 2915 beyond its star-forming core (i.e. radii of 42 -- 510 arcsec) to be $6.1\times 10^{-3}\,\rm M_{\sun}\,yr^{-1}$, assuming that the derived SFL holds out to the limits of the \HI\ data of \citet{Elson:2010iv} and a Kroupa IMF. This equates to a very low face-on intensity of $1.8\times 10^{-5}\,\rm M_{\sun}\,yr^{-1}\, kpc^{-2}$. Nevertheless, star formation in this region would be significant, equating to 12\% of the star formation of the inner region \citep[$\rm 4.9 \times 10^{-2}\, M_{\sun}\,yr^{-1}$ corrected to a Kroupa IMF; ][]{Werk:2010fn} or 11\% of the total SFR. The star formation in the outskirts of NGC 2915 implies significant metal production, which may explain the flat metallicity gradient observed by \citet{Werk:2010fn}. This assumes a Kroupa IMF, however, our results show that the IMF slope in the outer-disc region is steeper than a Kroupa IMF. If so, the outer disc will have a higher \SFI\ than stated. Adopting best-fitting IMF parameters of $\rm M_u=60,M_{\sun}$ and $\alpha=-2.85$ for $\rm M>1\, M_{\sun}$, and a Kroupa IMF for $\rm 0.08\, M_{\sun}<M<1\, M_{\sun}$, the \SFI\ will be a factor of 3.2 higher (Appendix A Table \ref{tab:ratios}). Since the centre of NGC 2915 is a high surface-brightness BCD \citep{Meurer:1994kv} and observations of high surface-brightness galaxies are consistent with a normal IMF \citep{Meurer:2009gp}, then the IMF may vary from a standard Kroupa IMF in the centre to one deficient in high-mass stars in the outer disc. If so, the SFR of the outer disc may be 40\% of the SFR in the centre, and comprise 28\% of the total SFR.

The extended \HI\ disc of NGC 2915 exhibits a constant velocity dispersion, $\sigma_{gas} \sim 10 \rm \, kms^{-1}$ \citep{Elson:2010iv}, which was puzzling because there appeared to be no star formation occurring beyond the central region of the galaxy, and thus no source to drive the observed turbulence \citep[e.g.][]{Wada:2002ep, Dib:2006dc}. We have now identified the likely source of turbulence as low density O and B star formation in the outer disc. These high-mass stars produce stellar winds and supernovae that provide the energy required for the observed velocity dispersion \citep{Dobbs:2011gp}. The existence of such stars is not surprising; star formation in outer discs is now thought to be common with the discovery of extended UV discs in $\sim$30\% of nearby galaxies \citep{Thilker:2007ho}.

\section{Conclusions}
\label{sec:con}
We have examined star formation in the outer \HI\ disc of the dark matter dominated, BCD galaxy NGC 2915. This environment, previously not known to harbour stars, is shown to have a low optical surface-brightness, low star formation intensity, and young populations of main-sequence disc stars. In addition to the clumpy distribution of main-sequence stars, which is the main subject of this paper, our CMD analysis shows a smooth distribution of RGB and ABG stars, which easily dominate the total number of NGC 2915 stars observed. We determine the nature of star formation in this environment using the main-sequence stars in the CMDs, which we then compare to simulations, along with published \Ha\ \citep{Werk:2010fn} and \HI\ \citep{Elson:2010iv} observations. 

We place constraints on the form of the IMF by comparing the main-sequence stars in the observed main-sequence luminosity function to simulations and \Ha\ observations, while assuming a constant star formation rate. The MS luminosity function analysis suggests than an IMF with a power-law slope $\alpha=-2.85$ and an upper-mass limit of $\rm M_u=60\, M_{\sun}$ is preferred. Using the main-sequence luminosity function analysis alone we are unable to constrain the upper-mass limit due to the insensitivity of optical colours to the masses of main-sequence stars.

We use previously published \Ha\ observations to place independent constraints on the upper-mass limit of the IMF. If we assume that all of the \Ha\ emission in the outer disc is confined to \HII\ regions then a single O star is required to produce to observed \Ha\ flux. In this case, the upper-mass limit of the IMF is restricted to $\rm M_u<20\, M_{\sun}$. The \Ha\ observations are at odds with the observed MS stars from the CMD and best-fitting IMF, which suggest there should be dozens of $\rm M_{\star}>20\, M_{\sun}$ present and an upper-mass limit $\rm M_u \sim 60\, M_{\sun}$. If all \Ha\ is confined to \HII\ regions then the objects we identify as high luminosity MS stars are likely to be binaries or compact clusters of lower-mass stars. This would then imply an IMF more extreme than our best-fitting IMF, having even less of the highest mass stars than a standard Kroupa IMF. On the other hand, if all of the observed stars in the MS luminosity function are single stars then we would have to discount the \Ha\ observations, implying a large fraction of escaping ionizing photons or diffuse \Ha. 

Using the same set of extensive simulated CMDs and the observed main-sequence stars we determine that the star formation law in the observed region follows a Kennicutt-Schmidt parametrization, with slope $\rm N=1.53 \pm 0.21$ and $\rm A=(9.6 \pm 3.3) \times 10^{-6}\, M_{\sun}\, yr^{-1}\, kpc^{-2}$, this is for an assumed Kroupa IMF in order to allow easy comparison with other results. If our best-fitting IMF is adopted, then A is a factor of 4.5 higher. Fits to this star formation law require a mass density of stars and dark matter on the order of $\rm 0.01\, M_{\sun}\, pc^{-3}$, which with our IMF results can be comprised of the known forming disc stellar populations, without requiring disc dark matter. If the observed star formation law holds to the edge of the observed \HI\ disc then star formation in the outer disc accounts for 11\% of the total SFR in the system if a uniform Kroupa IMF is assumed, and up to 28\% if the IMF is Kroupa in the centre and our best-fitting IMF over the extended \HI\ disc. Either way, star formation in the outer disc is a significant fraction of the total star formation in this isolated BCD galaxy.

\section*{Acknowledgements}

We would like to thank the referee for their helpful comments. We thank Tom Megeath, David Thilker, and Claus Leitherer for useful conversations. CL visited the University of Western Australia thanks to the visitor grant funded by the Centre for All-Sky Astrophysics (CAASTRO), which is an Australian Research Council Centre of Excellence (grant CE11E0090). This project was initiated as part of the Advanced Camera for Surveys (ACS) Instrument Definition Team effort. ACS was developed under NASA contract NAS 5-32865, and this research has been supported by NASA grant NAG5-7697 and by an equipment grant from Sun Microsystems, Inc. This research made use of: the NASA/IPAC Extragalactic Database (NED) which is operated by the Jet Propulsion Laboratory, California Institute of Technology, under contract with the National Aeronautics and Space Administration; NASA's Astrophysics Data System Bibliographic Services; and the VizieR catalogue access tool, CDS, Strasbourg, France. The original description of the VizieR service was published in A\&AS 143, 23.

\bibliography{papers_lib.new}

\section*{Appendix A}
\subsection*{Star formation rates from main-sequence stars}
We use the number of MS stars in the observed region of NGC 2915 to determine the star formation rate by comparing the number of observed MS stars to simulated stellar populations. We assume a constant SFR over the simulation time-scale $\rm \Delta t=600 \, Myr$, which is designed to be greater than the MS lifetime of the stars in the simulation. We determine the ratio, $\rm R_{sel}$, of the number of stars in the MS box, $\rm N_{sel}$, to the total number of stars formed over the observation time $\rm N_{\rm total}$ in the simulations. Hence:
\begin{equation}
\rm N_{total}=\frac{N_{obs}}{R_{sel}}.
\end{equation}

These simulations have a lower mass limit of $\rm M_{\star}=2.5\,M_{\sun}$ so the SFR for stars $\rm M_{\star}>2.5\,M_{\sun}$ is:
\begin{equation}
\rm \Psi^{\prime}_{sim}=\frac{N_{total}\overline{M^{\prime}_{\star}}}{\Delta t}.
\end{equation}
$\rm \overline{M^{\prime}_{\star}}$ is the average stellar mass for stars with $\rm M_{\star}>2.5\,M_{\sun}$:
\begin{equation}
\rm \overline{M^{\prime}_{\star}}=\frac{\int_{2.5}^{m_{u}} m\,\xi(m)\, \textrm{d}m}{\int_{2.5}^{m_{u}} \xi(m)\, \textrm{d}m}.
\end{equation}
The total SFR $\rm \Psi$ is then the ratio of the total mass formed in stars to the mass formed with $\rm M_{\star}>2.5_M{\sun}$:
\begin{equation}
\rm g_{2.5}=\frac{\int_{m_{l}}^{m_{u}} m\,\xi(m)\, \textrm{d}m}{\int_{2.5}^{m_{u}} m\,\xi(m)\, \textrm{d}m}.
\end{equation}
Where $\xi(m)$ is the IMF as defined in the introduction. Thus, the total SFR is given by:
\begin{equation}
\rm \label{eqn:SFR_2915_1}
\Psi=g_{2.5}\Psi^{\prime}=g_{2.5}\frac{N_{obs}}{R_{sel}} \frac{\overline{M^{\prime}_{\star}}}{\Delta t}.
\end{equation}
The inclination corrected number density per area of MS stars $\rm \Sigma_{\rm MS}$ is converted to the total star formation intensity, \SFI, using a modified version of Equation \ref{eqn:SFR_2915_1}:

\begin{equation}
\rm \label{eqn:SFR_2915_2}
\Sigma_{SFR}= g_{2.5}\frac{\Sigma_{MS}}{R_{sel}} \frac{\overline{M^{\prime}_{\star}}}{\Delta t},
\end{equation}
where $\Sigma_{MS}$ is the inclination corrected MS stellar density. Using Equation \ref{eqn:SFR_2915_2} the total $\Sigma_{SFR}$ of the observed field of NGC 2915 is $8.74\times10^{-5} \rm \,M_{\sun}\,yr^{-1}\,kpc^{-2}$ for a \citet{Kroupa:2001ki} IMF over a time-scale of 200 Myr, and the total SFR in this field is $2.45 \times 10^{-3}$ M$_{\sun}$yr$^{-1}$. The \SFI\ derived by \citet{Werk:2010fn} is $2.00 \times 10^{-4} \rm \,M_{\sun}\,yr^{-1}\,kpc^{-2}$ (corrected to a Kroupa IMF) for radii 44 -- 150 arcsec over the entire disc using the UV SFR conversion of \citet{Salim:2007hq} for sub-solar metallicity, and star formation averaged over a time-scale of 100 Myr. In comparison our observations in the ACS field cover radii 45 -- 257 arcsec for $\sim 1/6$ the of the outer disc over a time-scale of 200 Myr.  

The ratios $\rm g_{2.5}$ along with the $\rm \Sigma_{SFR}$ for various IMFs and are given in Table \ref{tab:ratios}. In each case the upper IMF slope ($\alpha$) extends down to $1\rm \,M_{\sun}$, and a Kroupa IMF is used for masses $0.08\, M_{\sun}<M<1\, \rm M_{\sun}$. $g_{2.5}$ is the ratio of the total mass formed in stars to the mass formed in stars with $\rm M_{\star} > 2.5\, M_{\sun}$, which increases with steeper $\alpha$, and decreases as $\rm M_u$ increases. The derived \SFI\ from MS stars has a weak dependence on $\rm M_u$ ($<5$\%) between $\rm M_u=120\, M_{\sun}$ and $\rm M_u=25\, M_{\sun}$ but a strong dependence on $\alpha$, up to a factor of $\sim 6$ over the range we explore. Our best-fitting IMF is a factor of $3.2$ heavier than a standard Kroupa IMF.

\begin{table}
\centering
\caption{Total SFR intensity calculated for different IMFs. In each case the IMF slope for $\rm 0.08\,M_{\sun}<M<1\, M_{\sun}$ is a \citet{Kroupa:2001ki} IMF and for $\rm M>1\,M_{\sun}$ the IMF slope is equal to $\alpha$.}
\label{tab:ratios}
\begin{tabular}{| c | c | c | c |c |}
\hline 
$\alpha$ & $\rm M_{u}$ & $\rm g_{2.5}$ & $\rm \Sigma_{SFR}$\\ 
 & ($\rm M_{\sun}$) & & ($\rm M_{\sun}\, yr^{-1}\, kpc^{-2}$) \\ \hline  \hline 
-2.35 & 120 & 3.5  & $8.74 \times 10^{-5}$ \\ 
-2.35 & 25 & 4.4 & $9.09 \times 10^{-5}$  \\
-2.85 & 120 & 9.4 & $28.9 \times 10^{-5}$ \\
-2.85 & 60 & 9.5 & $28.8 \times 10^{-5}$ \\
-2.85 & 25 & 10.4 & $30.4 \times 10^{-5}$ \\
-3.05 & 120 & 13.8 & $48.4 \times 10^{-5}$ \\
-3.05 & 25 & 14.8 &  $50.7 \times 10^{-5}$ \\ \hline
\end{tabular}
\end{table}

\label{lastpage}

\end{document}